\documentclass[twocolumn]{aastex631}

\newcommand{\bmodes}{{\it B}-modes}
\newcommand{\emodes}{{\it E}-modes}
\usepackage{graphicx}
\begin{document}
\title{\fontsize{18pt}{20pt}\selectfont Absolute reference for microwave polarization experiments.
\\
The COSMOCal project and its Proof of Concept.}

\correspondingauthor{Alessia Ritacco}
\email{alessia.ritacco@lpsc.in2p3.fr}
\author[0000-0003-0162-8206]{A. Ritacco}
\affiliation{Dipartimento di Fisica, Universit\`{a} di Roma ‘Tor Vergata’, via della Ricerca Scientifica 1, I-00133 Roma, Italy}\\
\affiliation{Laboratoire de Physique de l’$\acute{\rm E}$cole Normale Sup$\acute{\rm e}$rieure, ENS, Universit$\acute{\rm e}$ PSL, CNRS, Sorbonne Universit$\acute{\rm e}$, Universit$\acute{\rm e}$ de Paris, 75005 Paris, France}\\
\affiliation{INAF-Osservatorio Astronomico di Cagliari, Via della Scienza 5, 09047 Selargius, IT}\\
\affiliation{Univ. Grenoble Alpes, CNRS, LPSC-IN2P3, 53, avenue des Martyrs, 38000 Grenoble, France}
\author{L. Bizzarri}\affiliation{Dipartimento di Fisica, Universit\`{a} di Milano-Bicocca, Piazza della Scienza, 3, 20126, Milan, Italy} \affiliation{Laboratoire de Physique de l’$\acute{\rm E}$cole Normale Sup$\acute{\rm e}$rieure, ENS, Universit$\acute{\rm e}$ PSL, CNRS, Sorbonne Universit$\acute{\rm e}$, Universit$\acute{\rm e}$ de Paris, 75005 Paris, France}
\author{S. Savorgnano}\affiliation{Univ. Grenoble Alpes, CNRS, LPSC-IN2P3, 53, avenue des Martyrs, 38000 Grenoble, France}
      
\author[0000-0003-1097-6042]{F. Boulanger}\affiliation{Laboratoire de Physique de l’$\acute{\rm E}$cole Normale Sup$\acute{\rm e}$rieure, ENS, Universit$\acute{\rm e}$ PSL, CNRS, Sorbonne Universit$\acute{\rm e}$, Universit$\acute{\rm e}$ de Paris, 75005 Paris, France} 
\author{M. P\'{e}rault}\affiliation{Laboratoire de Physique de l’$\acute{\rm E}$cole Normale Sup$\acute{\rm e}$rieure, ENS, Universit$\acute{\rm e}$ PSL, CNRS, Sorbonne Universit$\acute{\rm e}$, Universit$\acute{\rm e}$ de Paris, 75005 Paris, France} 
\author{J. Treuttel}\affiliation{LERMA, Observatoire de Paris-PSL, 61 Avenue de l’Observatoire 75014
Paris France}
\author{P. Morfin}\affiliation{Laboratoire de Physique de l’$\acute{\rm E}$cole Normale Sup$\acute{\rm e}$rieure, ENS, Universit$\acute{\rm e}$ PSL, CNRS, Sorbonne Universit$\acute{\rm e}$, Universit$\acute{\rm e}$ de Paris, 75005 Paris, France}
\author{A. Catalano}\affiliation{Univ. Grenoble Alpes, CNRS, LPSC-IN2P3, 53, avenue des Martyrs, 38000 Grenoble, France}
\author{D. Darson}\affiliation{Laboratoire de Physique de l’$\acute{\rm E}$cole Normale Sup$\acute{\rm e}$rieure, ENS, Universit$\acute{\rm e}$ PSL, CNRS, Sorbonne Universit$\acute{\rm e}$, Universit$\acute{\rm e}$ de Paris, 75005 Paris, France}
\author{N. Ponthieu}\affiliation{Univ. Grenoble Alpes, CNRS, IPAG, 38000 Grenoble, France}
\author{A. Feret}\affiliation{LERMA, Observatoire de Paris-PSL, 61 Avenue de l’Observatoire 75014
Paris France}
\author{B. Maffei}\affiliation{IAS, CNRS, Universit\'{e} Paris-Saclay, CNRS, B\^{a}t. 121, 91405 Orsay, France}
\author{A. Chahadih}\affiliation{IAS, CNRS, Universit\'{e} Paris-Saclay, CNRS, B\^{a}t. 121, 91405 Orsay, France}
\author{G. Pisano}\affiliation{Dipartimento di Fisica, Sapienza Universit\`{a} di Roma, 00185 Roma, Italy}
\author{M. Zannoni}\affiliation{Dipartimento di Fisica, Universit\`{a} di Milano-Bicocca, Piazza della Scienza, 3, 20126, Milan, Italy}
\author{F. Nati}\affiliation{Dipartimento di Fisica, Universit\`{a} di Milano-Bicocca, Piazza della Scienza, 3, 20126, Milan, Italy}
\author{J. F. Mac\'{i}as-P\'{e}rez}
\affiliation{Univ. Grenoble Alpes, CNRS, LPSC-IN2P3, 53, avenue des Martyrs, 38000 Grenoble, France} 
\author{F. Cuttaia}\affiliation{INAF-Osservatorio di Astrofisica e scienza dello Spazio, Via Piero Gobetti, 93/3, 40129 Bologna BO}
\author{L. Terenzi}\affiliation{INAF-Osservatorio di Astrofisica e scienza dello Spazio, Via Piero Gobetti, 93/3, 40129 Bologna BO}
\author{A. Monfardini}\affiliation{Universit\'{e} Grenoble Alpes, CNRS, Institut N\'{e}el, France} 
\author{M. Calvo}\affiliation{Universit\'{e} Grenoble Alpes, CNRS, Institut N\'{e}el, France}
\author{M. Murgia}\affiliation{INAF-Osservatorio Astronomico di Cagliari, Via della Scienza 5, 09047 Selargius, IT}
\author{P. Ortu}\affiliation{INAF-Osservatorio Astronomico di Cagliari, Via della Scienza 5, 09047 Selargius, IT}
\author{T. Pisanu}\affiliation{INAF-Osservatorio Astronomico di Cagliari, Via della Scienza 5, 09047 Selargius, IT} 
\author{J. Aumont}\affiliation{Universit\'{e} de Toulouse, UPS-OMP, CNRS, IRAP, 31028 Toulouse, France} 
\author{J. Errard}\affiliation{Universit\'{e} Paris Cité, CNRS, Astroparticule et Cosmologie, F-75013 Paris, France} 
\author{S. Leclercq}\affiliation{Institut de Radioastronomie Millim\'{e}trique (IRAM), 38406 Saint Martin d’H\`{e}res, France} \author{M. Migliaccio}\affiliation{Dipartimento di Fisica, Universit\`{a} di Roma ‘Tor Vergata’, via della Ricerca Scientifica 1, I-00133 Roma, Italy}

\begin{abstract}
\textit{Context.} The cosmic microwave background (CMB), a remnant of the Big Bang, provides unparalleled insights into the primordial universe, its energy content, and the origin of cosmic structures. The success of forthcoming terrestrial and space experiments hinges on meticulously calibrated data. Specifically, the ability to achieve an absolute calibration of the polarization angles with a precision of $<0.1^\circ$ is crucial to identify the signatures of primordial gravitational waves and cosmic birefringence within the CMB polarization.
\textit{Aims.} We introduce the COSMOCal project, designed to deploy a polarized source in space for calibrating microwave frequency observations. The project aims to integrate microwave polarization observations from small and large telescopes, ground-based and in space, into a unified scale, enhancing the effectiveness of each observatory and allowing robust combination of data.
\textit{Methods.} To demonstrate the feasibility and confirm the observational approach of our project, we developed a prototype instrument that operates in the atmospheric window centered at 260~GHz, specifically tailored for use with the NIKA2 camera at the IRAM 30 m telescope.
\textit{Results.} We present the instrument components and their laboratory characterization. The results of tests performed with the fully assembled prototype using a KIDs-based instrument, similar concept of NIKA2, are also reported.
\textit{Conclusions.} This study paves the way for an observing campaign using the IRAM 30m telescope and contributes to the development of a space-based instrument.
\end{abstract}


\section{Introduction} \label{sec:intro}

\label{sec:introduction}
The Cosmic Microwave Background (CMB) radiation, which represents the faint afterglow of the Big Bang, offers a remarkable point of view of the early universe, providing profound insights into its structure, evolution, and fundamental cosmological parameters \citep{hu2002}. In recent decades, precision measurements of the CMB have significantly improved our understanding of the cosmos \citep{planck2018I}, validating the Big Bang theory and supporting the inflationary paradigm \citep{Planck2018X}. Furthermore, the polarization signal of the CMB unravels valuable information regarding energy scales, particle interactions, and the nature of primordial fluctuations, which later evolved into the cosmic structures observable today \citep{2002Natur.420..772K, 1968ApJ...153L...1R}.
The CMB radiation exhibits two primary polarization patterns: i) the \emodes~polarization, characterized by coherent and symmetrical signals originating mainly from density fluctuations in the early universe, and ii) the \bmodes~polarization, in which the electric field oscillations of the CMB photons display a curl-like nature, forming circular and spiral patterns \citep{HU1997323}.

The ongoing development of CMB experiments, including LiteBIRD \citep{LiteBIRD:2022cnt}, the Simons Observatory (SO) \citep{Simons19}, CMB-S4 \citep{2016arXiv161002743A}, and others, aims to achieve unprecedented levels of sensitivity in the detection of CMB polarization. These advancements hold the potential to investigate two crucial epochs in cosmic history: i) the reionization era when the first stars and galaxies emerged, leading to the ionization of neutral hydrogen throughout the universe \citep{10.1093/mnras/staa2797}; ii) the cosmic inflation epoch marked by rapid expansion just a fraction of a second after the universe's birth \citep{linde1982new}. 
Detecting inflationary gravitational waves, enclosed in the CMB \bmodes~polarization pattern \citep{polnarev1985polarization}, would directly confirm the occurrence of cosmic inflation, providing a pathway to explore the fundamental physics of the early universe that extends beyond the limitations of the Standard Model \citep{PhysRevD.23.347}. The energy scale of cosmic inflation is quantified by the tensor-to-scalar ratio $r$ that measures the amplitude of primordial \bmodes~relative to \emodes~ in the CMB. This parameter is predicted to range from 10$^{-2}$ to 10$^{-4}$ \citep{Kamionkowski16}.

The necessary sensitivity to achieve this detection represents a crucial advancement, which requires substantial improvements in both instruments and data analysis techniques compared to the past \citep{2021PhRvD.103f3507V}. 
One of the most critical aspects regards the absolute calibration of the detector's orientation. Any bias introduced by the experiment could significantly impact the control of systematic effects, the subtraction of galactic foreground emissions, and the minimization of the polarization $E$ to $B$ leakage. To date, self-calibration techniques have been used to mitigate instrumental errors and uncertainties \citep{Keating}. However, these techniques demand model assumptions on the nature of CMB polarization, preventing us from new discoveries in the field, such as cosmic birefringence \citep{2020PhRvL.125v1301M, 2023JCAP...01..044D,PhysRevD.108.082005}, or cosmic polarization rotation, which can be caused by parity violating extensions of the standard model \citep{1990PhRvD..41.1231C,PhysRevLett.103.051302} or primordial magnetic fields \citep{1990PhRvD..41.1231C,2022PhRvL.128i1302D}. %FB: reformulation
Furthermore, \cite{2023A&A...670A.163R} and \cite{2023A&A...672A.146V} have highlighted the difficulty of developing a robust model of Galactic polarization power spectra that takes into account the impact on the decomposition into \emodes~and \bmodes~of the coupling between the distribution of the spectral energy of dust emission and the orientation of the magnetic field.

To reduce reliance on model assumptions with regard to the characterization of the experiments, as well as models of the CMB and Galactic foregrounds, it is essential to have an independent means to calibrate polarization angles. Future experiments aim for a sensitivity that requires polarization angle measurements to be ten times more accurate than needed for the \textit{Planck} mission \citep{Rosset2010}. 
Addressing this challenge, the COSMOCal ({\bf COS}mological {\bf M}icrowave {\bf O}bservations {\bf Cal}ibrator) project seeks to establish a method for the calibration of polarization angles with ground-based telescopes, facilitating the comparison of data among telescopes with apertures ranging from small ($\sim 40\,$cm) to large ($> 5\,$m).

The article is organized as follows. 
Sec.~\ref{sec:the idea} presents the COSMOCal project, outlining its scientific rationale and the overall framework.
Sec.~\ref{sec:the prototype} introduces a prototype device customized for use with the NIKA2 camera at the IRAM 30 m telescope. The millimeter source and the optical system designed to measure the polarization orientation are presented with their laboratory characterization in Secs.~\ref{sec:the millimeter source} and \ref{sec:optical design}. 
Laboratory measurements with the fully assembled instrument,  using a millimeter camera with KIDs (Kinetic Inductance Detectors) detectors are reported in Sec.~\ref{sec:lab_tests_grenoble}.
Sec.~\ref{sec:conclusions} presents the conclusions and future plans to perform tests with the IRAM $30\,$m telescope and design a space instrument.

\begin{figure*}[htb!]
    \centering
    \includegraphics[width=1.3\columnwidth]{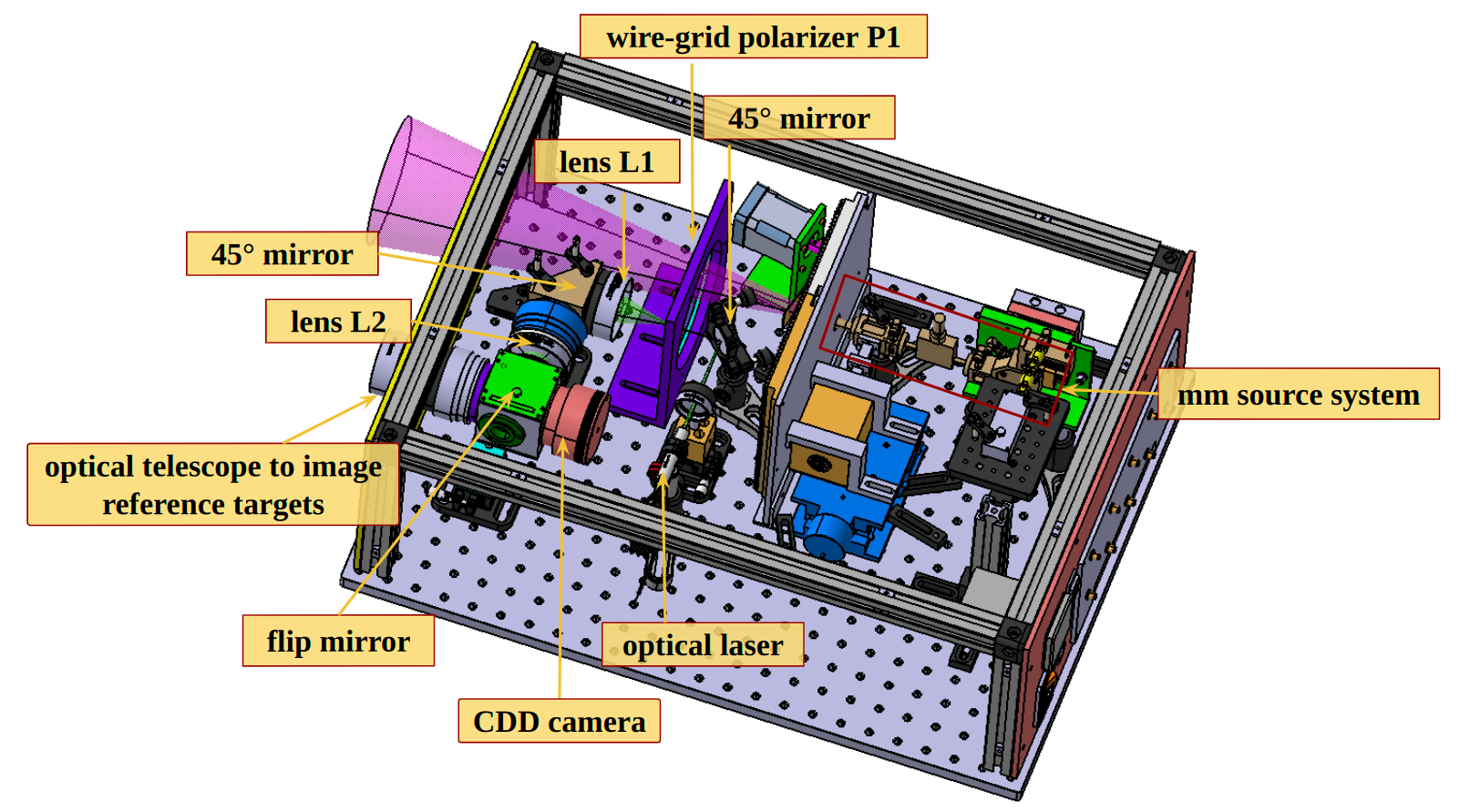}
    \includegraphics[width=0.6\columnwidth]{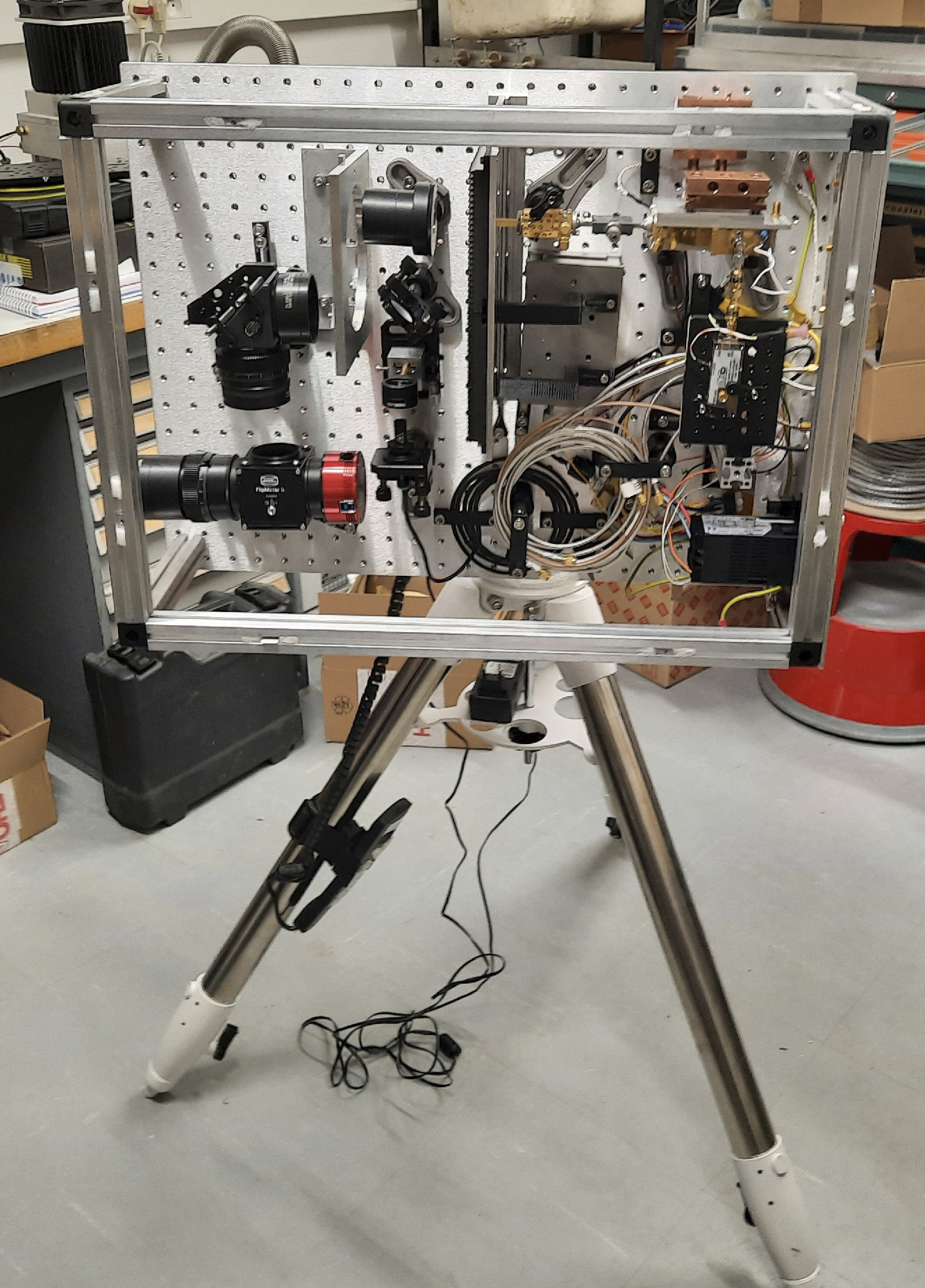}
    \caption{{\it Left}: mechanical drawings of the COSMOCal box containing the: microwave source components (right) and the optical system (left). {\it Right}: first assembled version.}
    \label{box inside}
\end{figure*}

\section{COSMOCal project}
\label{sec:the idea}

\subsection{Project rationale}

To analyze results from upcoming CMB experiments, independently of the standard cosmological model and Galactic foregrounds, we propose developing a calibration method that does not involve self-calibration on the microwave sky emission. This initiative is currently being implemented using ground-based or drone-mounted sources \citep{Nati17, coppi2022, Cornelison22}, and developing a cube-sat to be placed into the second Lagrange point L2 in orbit around a CMB telescope as proposed by \cite{2021Senso..21.3361C}. However, these efforts only apply to the smallest ground based CMB telescopes or a specific space mission. 
To calibrate ground-based large telescopes, the source must be in space to be in their far field. The easiest step into space would be to launch the source on a nano-satellite into a low-Earth orbit as originally proposed by \citet{Johnson2015}. However, in such an orbit the source would move too quickly across the sky to be tracked by large aperture microwave telescopes on Earth. The alternative we are considering for COSMOCal is to place the source as a guest payload on a satellite in geostationary orbit. In addition, having a reference source at a fixed point in the sky allows us to establish a reproducible calibration strategy with ground-based telescopes across several years. 

The COSMOCal project is set to pioneer a novel method for calibrating CMB observations through a space experiment designed to achieve the accuracy necessary to fulfill exceptional scientific goals.
We aim to build a multi-frequency source to be deployed in space, which emits a polarization signal with a highly precise orientation of $< 0.1 ^{\circ}$ within the frequency range of 90-300 GHz. 
 The COSMOCal source will help integrate microwave polarization observations from terrestrial and space-based platforms, focused on exploring the CMB and Galactic astrophysics, on a unified scale that maximizes the effectiveness of each observatory and facilitates a powerful combination of data from small and large telescopes. 
 
 The LiteBIRD space experiment and telescopes at the South Pole would not be able to observe the COSMOCal source, but once large ground-based telescopes are calibrated to the required accuracy, they may be used to observe a list of astrophysical polarized sources, such as the Crab supernova remnant, to the required precision. These astrophysical sources will become calibration standards shared by all telescopes, on the ground and in space. 

\subsection{Project framework}

The development of our project is divided into three main steps. 
First, to establish a proof-of-concept and validate the observational strategy that underlies the COSMOCal project, we constructed a prototype instrument designed to operate within the atmospheric window centered at 260 GHz. The presentation of this instrument with its laboratory characterization is the main purpose of this article. The second step will be the design and study of a multi-frequency space instrument that emits a polarization signal strong enough to be detected by large telescopes on Earth. The last is the deployment of a payload with the calibration source in space, in a geostationary orbit. 

The COSMOCal framework involves the use of several ground-based telescopes. This article introduces a prototype device intended for operation at 260 GHz at the IRAM $30\,$m telescope. The source will be placed on the top of the Pico Veleta a few kilometers from the telescope. Observations of the COSMOCal source will be coordinated with observations on reference astrophysical sources, such as the Crab Nebula. The Crab Nebula currently serves as the primary sky calibrator for polarization observations at IRAM, but its observations are hindered by instrumental calibration uncertainties of about 1$^\circ$ \citep{2018A&A...616A..35R,2020A&A...634A.100A}. 
The COSMOCal project aims to greatly improve this accuracy and provide regular monitoring of this source. 

In addition to the Crab Nebula, other sources will be used as secondary calibrators, further contributing to the project goals. Plans are underway to incorporate a 90 GHz source in the prototype for a test campaign at the Sardinia Radio Telescope as well.

The space instrument will be positioned at longitude on the Earth's equator, allowing observations from telescopes located in southern Europe and Chile, such as the Simons Observatory \citep{2019JCAP...02..056A}, which started operations in 2024. 
The calibration of a selected set of large telescopes will be the first scientific outcome of COSMOCal. The second one will be Stokes Q and U maps of the reference astrophysical sources. These maps will become references for use to calibrate microwave polarization observations of all telescopes, including the telescopes at the South Pole and LiteBIRD in space, which cannot directly calibrate their observations observing the COSMOCal source.

To fulfill the ambitious goals of the COSMOCal project, significant hurdles in data analysis must be addressed because of the integration of polarization angle measurements within the overall data calibration process. The necessary calibration precision requires the identification and correction of instrumental effects that affect the observations. This includes characterizing the polarization beam of the telescope, evaluating the leakage from intensity to polarization, and reducing cross-polarization. Moreover, advanced data analysis methods are essential to effectively combine sky maps from telescopes that have very different angular resolutions.

\section{Prototype instrument}
\label{sec:the prototype}
To establish a proof-of-concept and validate the observational strategy that underlies the COSMOCal project, we constructed a prototype instrument at a frequency to operate within the atmospheric window centered at 260 GHz. This instrument is designed for a test campaign at the IRAM 30m telescope with the NIKA2 camera \citep{2020A&A...637A..71P}.

The NIKA2 polarization system (a.k.a NIKA2pol) showcases remarkable sensitivity of 20 mJy.s$^{1/2}$ in identifying sky polarization \citep{nika2pol}; thanks to the continuously rotating half-wave-plate that shifts the signal at higher frequencies in Fourier space, effectively separating it from low-frequency noise \citep{2017A&A...599A..34R}.
However, the uncertainty associated with polarization angle measurements is so far limited to  1$^\circ$ \citep{2022EPJWC.25700042R}, which was determined during the NIKA2pol commissioning phase.
\newline
\newline
The current design of the COSMOCal prototype is shown in Fig.~\ref{box inside}. The instrument, housed in a thermally insulated box, consists of the following components.
\begin{itemize}
\item 
A radio frequency chain that generates a monochromatic signal at 265 GHz and directs it through a waveguide to a 16$^\circ$ beam horn,
\item
 A metal grid polarizer (P1) that ensures the purity of the polarized signal,
 \item
An optical laser used to generate a diffraction pattern created by the grid of wires in the polarizer P1,
\item 
Optics to focus the image of the diffraction pattern onto a CDD camera, 
\item 
A flip mirror to alternatively image ground landmarks onto the CCD camera for determining the 3D position of the COSMOCal source by photogrammetry.
\end{itemize}

The diffraction pattern image makes it possible to recover and monitor the alignment of the wires, and thereby the orientation of the signal polarization.
%The millimeter source consists of a chain of radio-frequency (RF) components, while the optical system consists of various optical elements necessary to obtain an independent measurement of the polarization angle during observations.

The objective of the foreseen test campaign with NIKA2 is to evaluate the effectiveness of our technological concept and study all potential contributions through the optical chain to the final uncertainty associated to the measured polarization angle. This will serve as a test bench for future development of a space payload, already providing major insights into the precision of polarization angle reconstruction with the NIKA2 camera.
\\
\\
\subsection{The millimetre source}
\label{sec:the millimeter source}

\begin{figure*}[htb!]
\centering
\includegraphics[width=0.9\textwidth]{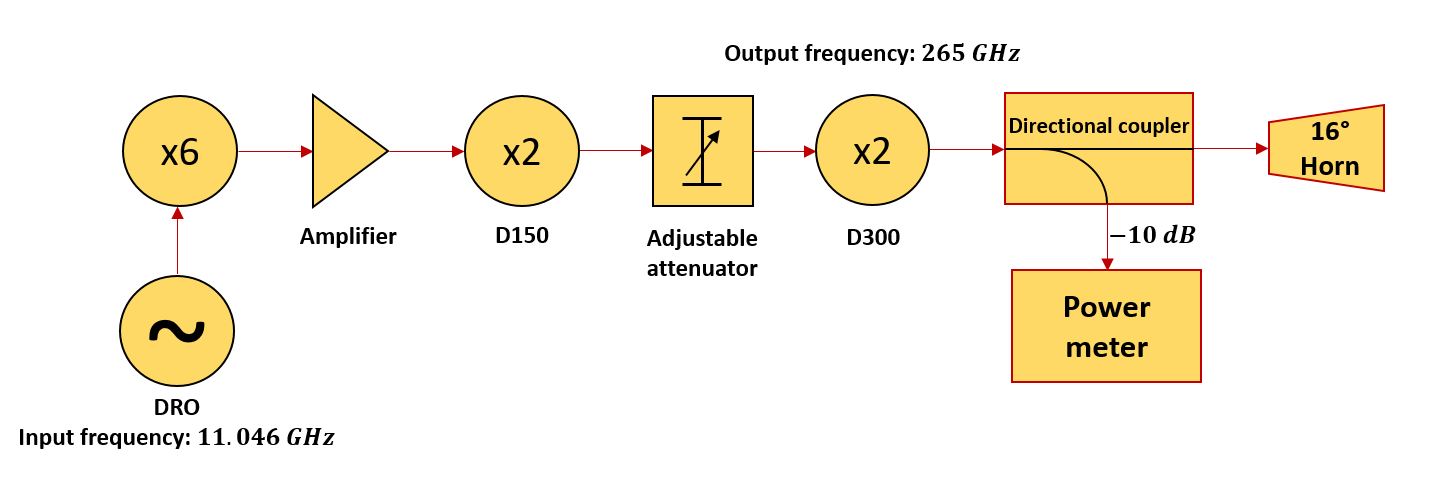}
\caption{COSMOCal radio-frequency chain scheme.}
\label{fig:microwave_schematic}
\end{figure*}

\begin{figure}
    \centering
    \includegraphics[width=\columnwidth]{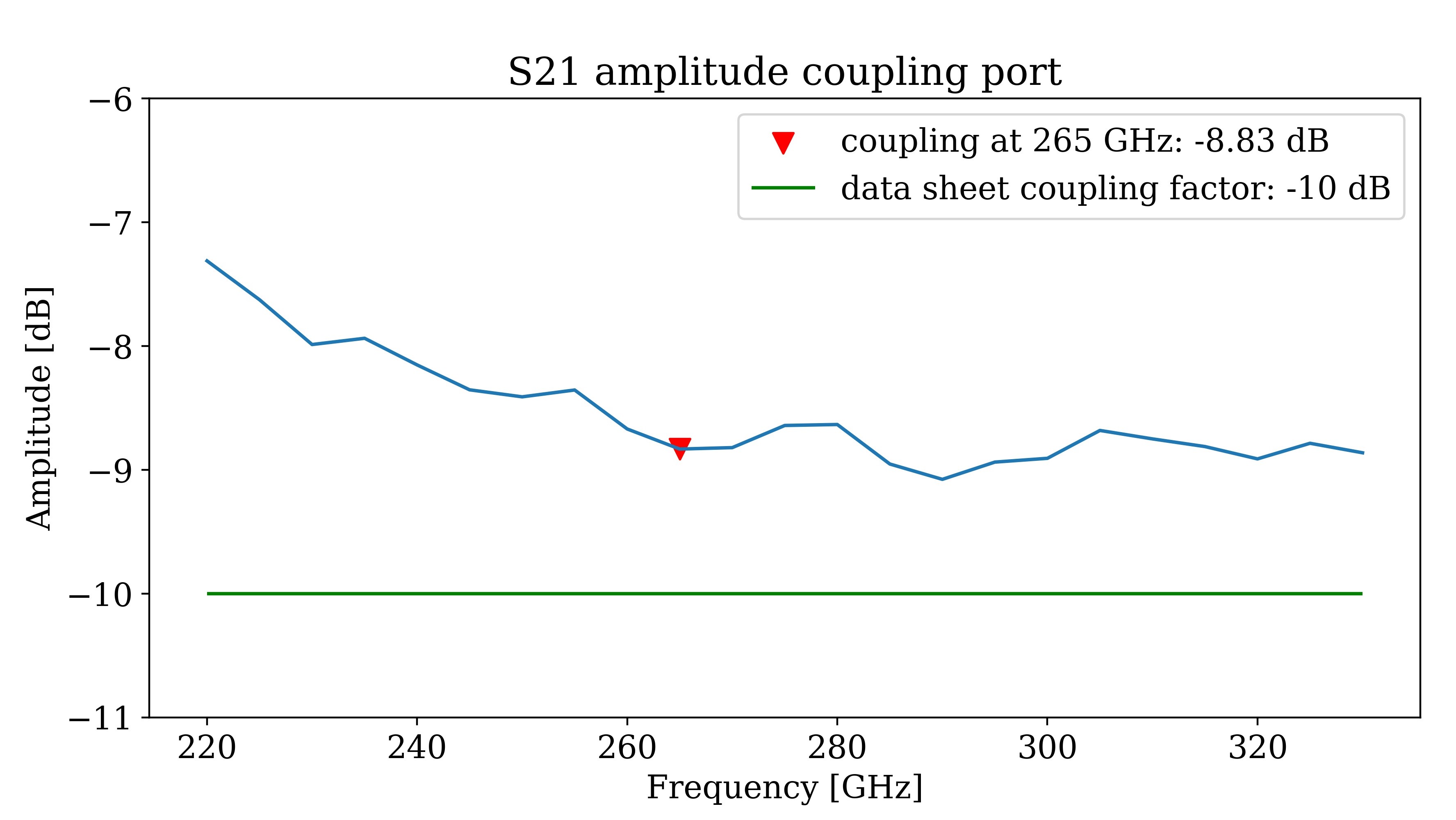}
    \caption{S21 parameter of the VNA representing the transmitted power (in dB) as measured at the output of the directional coupler's port. The coupling factor at 265 GHz is shown by a red triangle, as well as the value provided in the data sheet represented with a green line.}
    \label{fig:coupling_power}
\end{figure}

\begin{figure}
    \centering
    \includegraphics[width=\columnwidth]{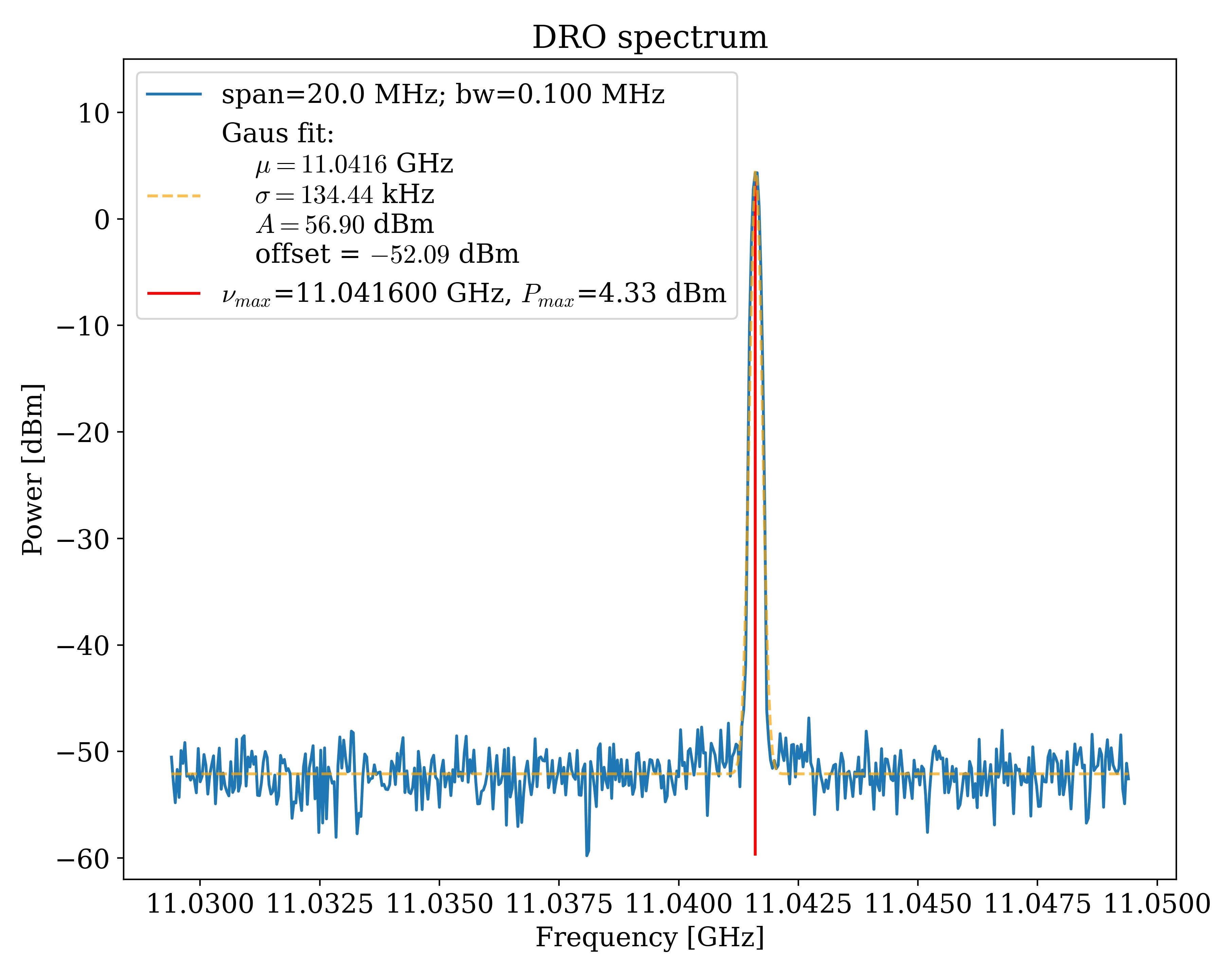}
    \caption{Measured spectrum of the DRO through a spectrum analyzer. The yellow dashed line represents a Gaussian fit performed on the spectrum, while the red line highlights the central frequency of the spectrum.}
    \label{fig:DRO_spectrum}
\end{figure}

The millimeter source, assembled at the LERMA institute, operates within a frequency range of approximately 260 to 310 GHz to align with the 1 mm frequency band of NIKA2 \citep{2022A&A...658A..24P}. As depicted in Fig.~\ref{fig:microwave_schematic}, the millimeter source chain employs a 24 times frequency multiplication scheme to emit a Gaussian beam monochromatic signal at 265 GHz. The key technologies used in the schematic have been space-qualified to TRL8 \citep{Treuttel}. The chain comprises a dielectric resonator oscillator (DRO) injecting a signal at 11.041 GHz into a W band sextupler (AFM 60-90), a W band power amplifier (E-MPA 66-80), two frequency doublers at 150 and 300 GHz, and one adjustable attenuator securing the interface ports matching and isolation. The output signal is coupled, to free space, with a 20 dBi pyramidal horn and monitored with a PM5 power meter through a 10 dB directional coupler located before the feed-horn. This monitoring of the proper functioning of the microwave source will also be ensured during the observing campaign at the IRAM 30m site, in addition to the frequency multipliers that rectify currents. The characterization of the coupler at different frequencies has been performed at the IAS institute using a vector network analyzer (VNA). The VNA's transmitted power measured at the output of the directional coupler port is shown in Fig.~\ref{fig:coupling_power}. The red triangle highlights the coupling factor as measured at 265 GHz. 
The frequency stability is given by the DRO 1000 generating an output signal at 11.0416 GHz with a power of 21 dBm, easily attenuated, and very stable over a temperature range of 0$^\circ$C to 50$^\circ$C.  The DRO spectrum is peaked around the central frequency, within a range $\sim$ 100 kHz, a phase noise of -97 dBc / Hz at 10 kHz, and a frequency dependence on temperature of 4.4 kHz/$^{\circ}$C, as illustrated in Fig.~\ref{fig:DRO_spectrum}.
%The yellow line corresponds to the Gaussian fit, yielding a peak frequency width of 134.44 kHz. 
This configuration meets the frequency stability requirements.

\subsection{Power constraints}
\label{sec:power constraints}

The COSMOCal calibration source will be positioned within the IRAM 30m telescope's field of view on the peak of Pico Veleta mountain, approximately 3 km away from the radio antenna. The determination of the required power to correctly illuminate the IRAM 30m telescope's focal plane and to avoid saturating the NIKA2's detectors is a crucial aspect. 
\newline
To this scope, we consider reasonable referring to the NIKA2's noise equivalent flux density (NEFD) measured in polarized intensity, of 20 $\pm$ 2 mJy/$\sqrt{\text{Hz}}$, in the $1$~mm frequency band \citep{nika2pol}. This flux sensitivity refers to a single KID, since for NIKA2 a single detector occupies the size of a beam. 
For convenience, the NEFD can be converted in a noise equivalent power (NEP) according to the following relation:
\begin{equation}
    \text{NEP} = A^\prime \times \, \Delta \nu \times \, \text{NEFD} \times{10^{-26}} 
\end{equation}
where $A^\prime$ is the collecting surface of the total detection system, $\Delta \nu$ is the frequency band of NIKA2, and $10^{-26}$ is a conversion factor.
The NIKA2 collecting surface is computed accounting for its cold pupil, being actually $A^\prime = 590\text{ m}^2$ of the IRAM telescope's primary mirror, while the frequency bandwidth of NIKA2 1 mm channel is: $\Delta \nu (\text{NIKA2}) \simeq 80$ GHz. The resulting NEP is therefore:
\begin{equation}
    \text{NEP}_{\text{NIKA2}} \simeq 9\times10^{-17} \text{W}\sqrt{\text{s}}
\end{equation}
This NEP corresponds to a noise equivalent temperature (NET) of $\sim 2
 \text{mK}\sqrt{\text{s}}$.
Therefore, it seems reasonable to require a received power of 10~pW on each NIKA2 detector, resulting in a signal-to-noise ratio (SNR) of approximately 200 considering the sampling frequency of NIKA2 for polarization measurements $\sim 47$~Hz, and ensuring a linear response of the detectors. Translating this result in temperature terms, a 10~pW power variation would correspond to a $\sim4.5$ K temperature variation.  NIKA2 detectors are Kinetic Inductance Detectors (KIDs) with a typical response of $\sim\frac{1\text{kHz}}{\text{K}}$, and a characteristic width of the resonance frequency of the order of $\sim$ 100 kHz. More details on the operation of this type of detectors are provided in Sec.~\ref{sec:lab_tests_grenoble}. This means that a $4.5$K signal would produce a frequency response of $\Delta\nu_{signal}\sim 4.5$ kHz, which is $\sim 5 \%$ of the typical resonance frequency width. This ensures the linearity of the electronic response to the resonance frequency variation of KIDs.

In the calculation of the received power for KIDs detectors, we have considered the impact of geometrical losses, which serve as the primary attenuation factor. Although the source's beam fully illuminates the primary mirror, only a small fraction is captured due to the field of view (FoV) limitation of the IRAM telescope, which spans 6.5 arc-minutes. Of the total $16^\circ$ beam, only $0.108^\circ$ reaches the detectors. Specifically, considering the placement of the calibrator at a distance of approximately 3 km from the 30 m antenna, the emitted power from the source experiences a dilution by a factor of approximately $\varepsilon_{geom}\simeq 2\times 10^4$.

In conjunction with these geometrical considerations, it is essential to incorporate an additional factor to address losses resulting from the approximately 35\% efficiency of the optics in transmitting radiation to the detector focal plane \citep{2018A&A...609A.115A}. Assuming uniform illumination across the detector's focal plane, we can derive a power of 10 nW at the detector focal plane. This establishes a requirement of approximately 1 mW for the power emitted by the millimeter source.
Moreover, to validate the assumption of focal plane uniformity, we used the Zemax optical software. In the simulations, we modeled a point source emitting a Gaussian beam with a 16$^{\circ}$ point spread function (PSF), positioned 3 km away from the telescope and including the IRAM telescope's optics. The results demonstrated an extended illumination of the focal plane, a zone being in the shadow of the secondary mirror and the quadrupod, which position will depend on the alignment of the source.

\subsubsection{Power stability}
\begin{figure}
    \centering
    \includegraphics[width=\columnwidth]{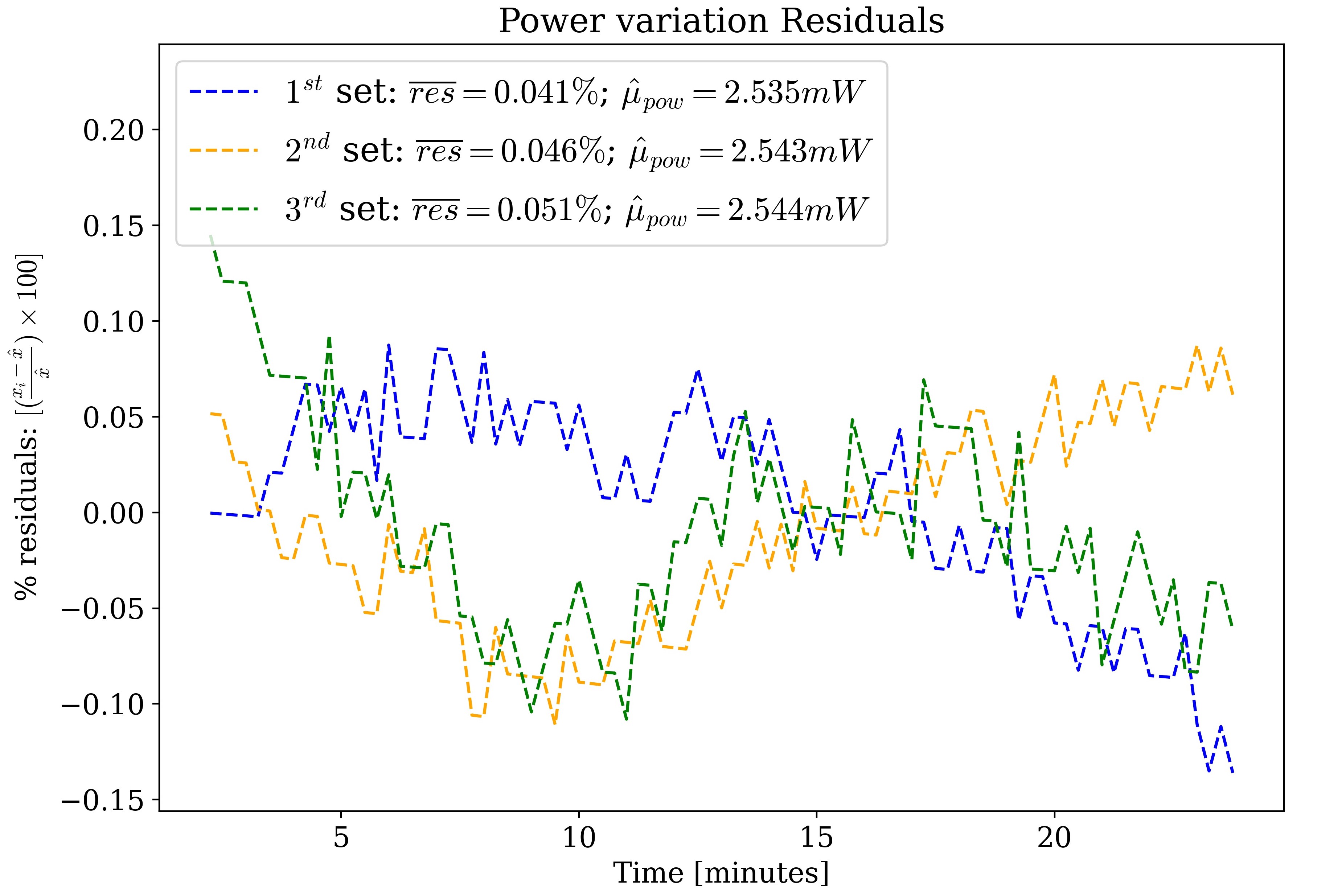}
    \caption{The three dashed lines illustrate the computed percent residuals from the average power of the source during a single measurement. The legend also displays the average residuals and total power.}
    \label{power stability}
\end{figure}

Several constraints shape the observation process. Maintaining consistent power throughout the NIKA2 observing sessions is a primary concern. The precision of polarization angle measurement, based on Stokes parameters measured by the NIKA2 camera $Q$ and $U$, is directly affected by measurement uncertainties. Thus, maintaining the emission stability of the artificial source throughout the observation is of critical importance.
Moreover, meeting the modulation conditions for the signal amplitude is essential. Directed towards the source location on the mountain, we require a signal amplitude modulation at frequency $\nu < 1$ Hz to effectively filtering out the background signal contamination during subsequent data analysis.\newline
To verify the power stability of the source we performed laboratory measurements for periods of 25-minutes in different moments of the day.
We quantified the oscillation of power during a single measurement by analyzing the percent residuals of the average value, as shown in Fig.~\ref{power stability}. Furthermore, Tab.~\ref{tab: pow residuals} presents the results from three sets of measurements of percent residuals over 5, 10, and 20-minute intervals.

%Here we can notice that the typical oscillation of the source's power during a single measurement is of the order of $1\mu$W.
\begin{table}[]
    \centering
    \begin{tabular}{|c|c|c|c|}
    \hline
     & $1^{st}$ set &  $2^{nd}$ set & $3^{rd}$ set \\
    \hline
    $res$ ($5$ min)* & 0.0412 $\%$ & 0.0468 $\%$ & 0.0529 $\%$ \\
    $res$ ($10$ min)* & 0.0413 $\%$ & 0.0467 $\%$   & 0.0538 $\%$ \\
    $res$ ($20$ min) & 0.0409 $\%$ & 0.0461 $\%$  & 0.0533 $\%$ \\
    \hline
    \end{tabular}
    \caption{The percentage residuals representing the deviation from the average value of the total source power measured at different times.
    \newline
    * Residuals for 5 and 10-minute intervals are calculated as averages across different time intervals of the same duration. }
    \label{tab: pow residuals}
\end{table}

As a conclusion, during typical NIKA2 measurements lasting around 10 minutes, power fluctuations are roughly $10^{-3}$ of the total output power, which makes them unimportant.

\subsection{Optical system: polarization angle accuracy}
\label{sec:optical design}

The optical system is shown on the left side of the calibrator's box in Fig.~\ref{box inside}. The goal of the COSMOCal optical system is to ensure the online determination of the polarization angle of the COSMOCal system with an accuracy of $\Delta\psi < 0.1^\circ$. \newline
Although the wave-guide in the RF chain naturally polarizes the signal, we lack a method to determine its orientation at the required precision and continuously monitor it during the observations. To address this difficulty, we implemented a polarizing grid (hereafter P1) in the optical system ensuring a pure polarization of the millimeter source's signal. In order to determine the orientation of the output polarization signal, we also include: a 520 nm optical laser; a 60 mm focal lens (L1), a flat 45$^\circ$ mirror, an helical focuser, a $100$ mm focal lens (L2), a $70$ mm f/6 refractor telescope, a flip mirror and a CCD camera. 
The laser shines through the polarizer P1 producing a diffraction pattern image that is detected by the CCD camera, similar to the strategy adopted by \cite{coppi2022} and \cite{Dunner2021}.
Then, the optical system works as follows: lens L1 shifts the diffraction pattern to infinity, and lens L2 (along with the helical focuser) concentrates the image onto the CCD camera. The flip mirror is used to alternate between the diffraction pattern and some ground's references images onto the CCD camera, providing continuous updates about their relative angle.\newline
The diffraction pattern's analysis enables us to determine the orientation of the polarized signal in the CCD camera plane. The ground references enable us to establish the three Euler angles of the camera with respect to the ENU (East, North, Up) coordinate system of the NIKA2 receiver cabin. In particular, the roll angle is measured around the line of sight and is needed to determine the orientation of the polarization as seen by the NIKA2 detectors (more details in Sec.~\ref{sec:Photogrammetry}).

The COSMOCal polarization uncertainty is mostly given by the accuracy on the diffraction pattern orientation and the roll angle reconstruction:
\begin{equation}
    \sigma_{\text{cal}} = \sqrt{\sigma_{\text{diffraction}}^2 + \sigma_{\text{roll}}^2}
\end{equation}
The uncertainty on the roll angle is affected by the accuracy on the positioning of the targets, and on the target identification algorithm.\newline

Accurately determining the angle of the polarization orientation in the camera plane, to within the strict requirement of $0.1^\circ$, is of utmost importance. In the upcoming subsection, we detail the characteristics of the diffraction pattern and introduce the algorithm we have developed to analyze it, providing the output COSMOCal polarization angle.

\subsubsection{Diffraction pattern analysis} 
\label{sec:Diffraction pattern analysis}

The laser ($\lambda=520$ nm) creates a diffraction pattern through the polarizing grid. The diffraction is produced by the very thin wires, and the observed pattern is related to the distance between the wires of $\sim10$ $\mu$m.  

\begin{figure}
    \centering
    \includegraphics[width=\columnwidth]{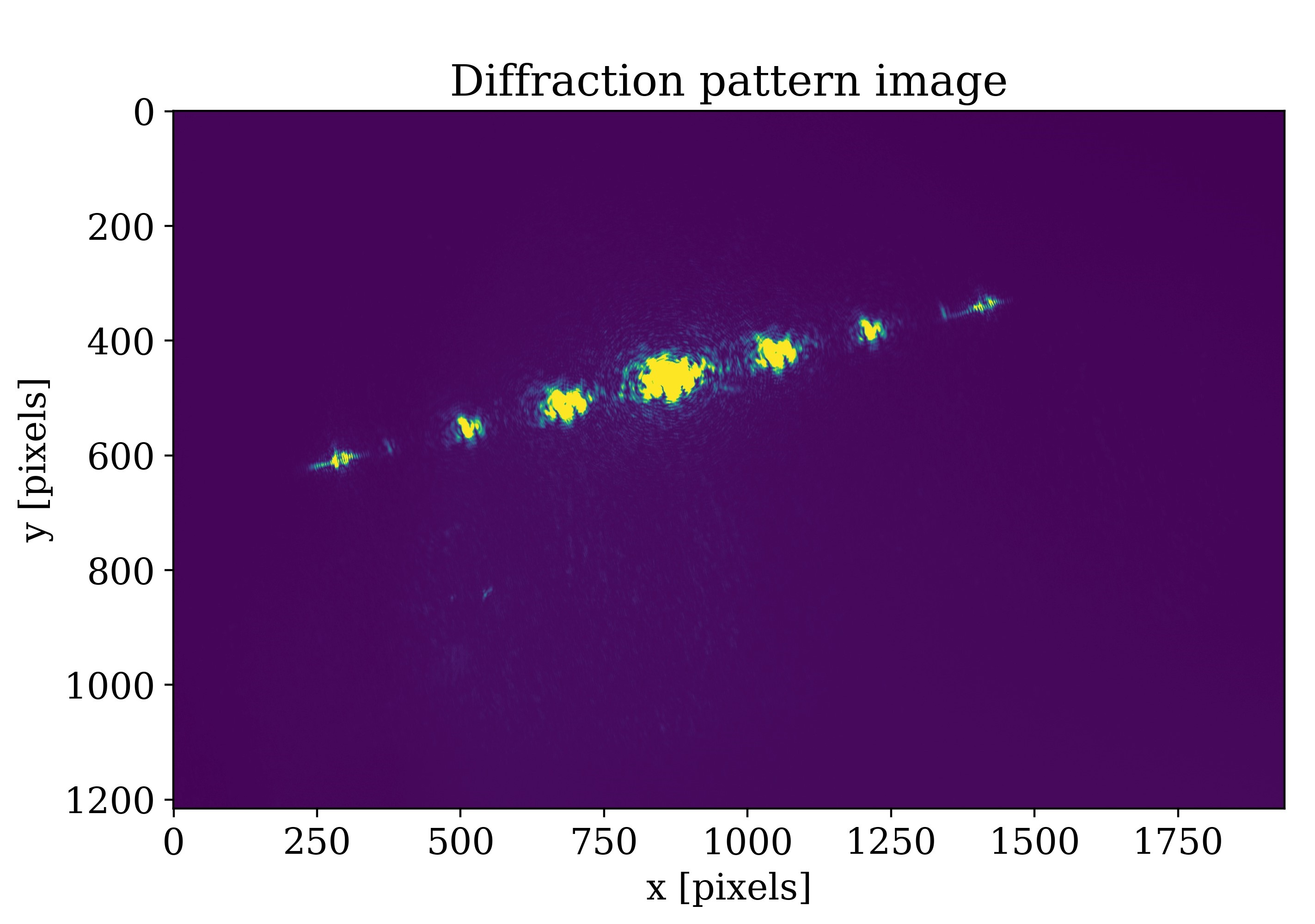}
    \caption{Typical image of the diffraction pattern acquired with the CCD camera. Image taken during the preliminary tests performed on the optical system.}
    \label{fig:diffraction pattern init image}
\end{figure}

\begin{figure}
    \centering
    \includegraphics[width=0.7\columnwidth]{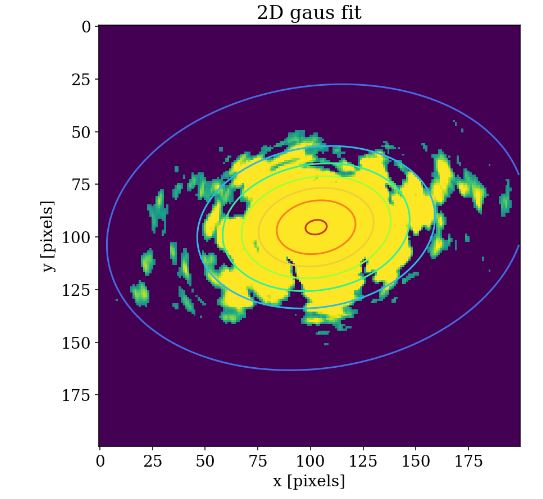}
    \caption{2D Gaussian fit performed on the neighborhood of a diffraction maximum.}
    \label{fig:2D gaus fit}
\end{figure}

The shape of the diffraction pattern can be approximated as a rotated parabola with two angles: $\varphi_i$ and $\theta$. \newline
$\varphi_i$ is the incident angle between the laser and the polarizer when the wires are parallel to the inclination of the polarizer, while $\theta$ gives the orientation of the polarization with respect to the $x$ axis of the CCD camera. If the incident angle $\varphi_i$ increases, the curvature of the parabola increases, while a rotation of the polarizer around its horizontal axis results in a variation of $\theta$.\newline
In Fig.~\ref{fig:diffraction pattern init image} a typical image of the diffraction pattern acquired by the CCD camera is shown.
The image is quite noisy, therefore we applied a  preliminary analysis to clean the image and have a better information on the maxima of diffraction. In particular, we set to zero each pixel that presents an intensity below 1.5 $\sigma$ from the intensity peak, where $\sigma$ is given by the Poisson error: $\sigma = \sqrt{S}$ with S being the signal amplitude. 
The data analysis includes the identification of the centroids of the maxima of diffraction, and the fit with a rotated parabola to retrieve the orientation of the polarization. \newline 
To identify the centroids we use a 2D Gaussian fitting described  by $7$ parameters: the amplitude $A$, the centroids $x_c$ and $y_c$, the spreads $\sigma_x$ and $\sigma_y$, the orientation in the camera plane $\theta_G$ and an offset C. Note that $\theta_G$ does not represent the angle that provides the polarization orientation. This is due to the fact that each centroid can exhibit a distinct orientation in the camera plane, which is derived from the curvature of the parabola.\newline
A typical result of the 2D Gaussian fit is shown in Fig.~\ref{fig:2D gaus fit}.
All the centroids identified in this way have an associated uncertainty of $\sim0.1$ pixels, indicating a good reliability of the method. \newline
With the centroids identified, we fit the diffraction pattern image with a rotated parabola function described by four parameters: the coordinates of the vertex: $(x_V,y_V)$, the curvature $a$ and the rotation angle $\theta$. In particular, $\theta$ gives us the relative angle between the orientation of the polarization and the $x$ axis of the camera. \newline

The fit is performed by using the \textit{emcee} algorithm \citep{2013PASP..125..306F}. We have developed two fitting procedures, differing one from the other depending on the assumptions made. \newline
The first fitting procedure assumed to know the parabola's vertex $(x_V,y_V)$, given by the coordinates of the centroid of the main diffraction maximum. In this case, we have only $a$ and $\theta$ as free parameters, and we applied flat priors to them, constraining $\theta$ within $[-90^\circ,90^\circ]$. 

In the second fitting procedure, we do not make any assumption on the vertex's position, leaving $x_V$ and $y_V$ as free parameters. In this case, we applied Gaussian prior to them, using the results of the 2D Gaussian fit performed on the main maximum of diffraction. \newline

We have tested these algorithms on several images of the diffraction pattern keeping the same orientation of the polarizer, in order to get information on the statistical error coming from the data analysis. 
\newline
Furthermore, in order to study any additional uncertainty that can occur from COSMOCal optics misalignment we have performed some tests by tilting different components of the optical system and checked the impact through the reconstruction of the diffraction pattern. 
In Tab.~\ref{tab:sigma results} the results of the error analysis on $\theta$ are reported, performed on ten different images acquired with the described procedure, and fitted with the two different methods. The total uncertainty on the orientation of the diffraction pattern in the camera plane is dominated by the statistical error (i.e. the standard deviation of the angles obtained through the data analysis), which is almost the same for the two fitting methods. Instead, the average systematic error computed by tilting parts of the optical system varies by almost an order of magnitude between the two fitting methods, but remains subdominant compared to the statistical error. 
Although the total uncertainty in $\theta$ is below the requirement of $0.1^\circ$ in both cases, we decide to favor the second fitting method over the first one. Indeed, it concedes more freedom in the definition of the parabola's vertex, without losing accuracy in the final estimation of the $\theta$ parameter.

\begin{figure}
    \centering
    \includegraphics[width=\columnwidth]{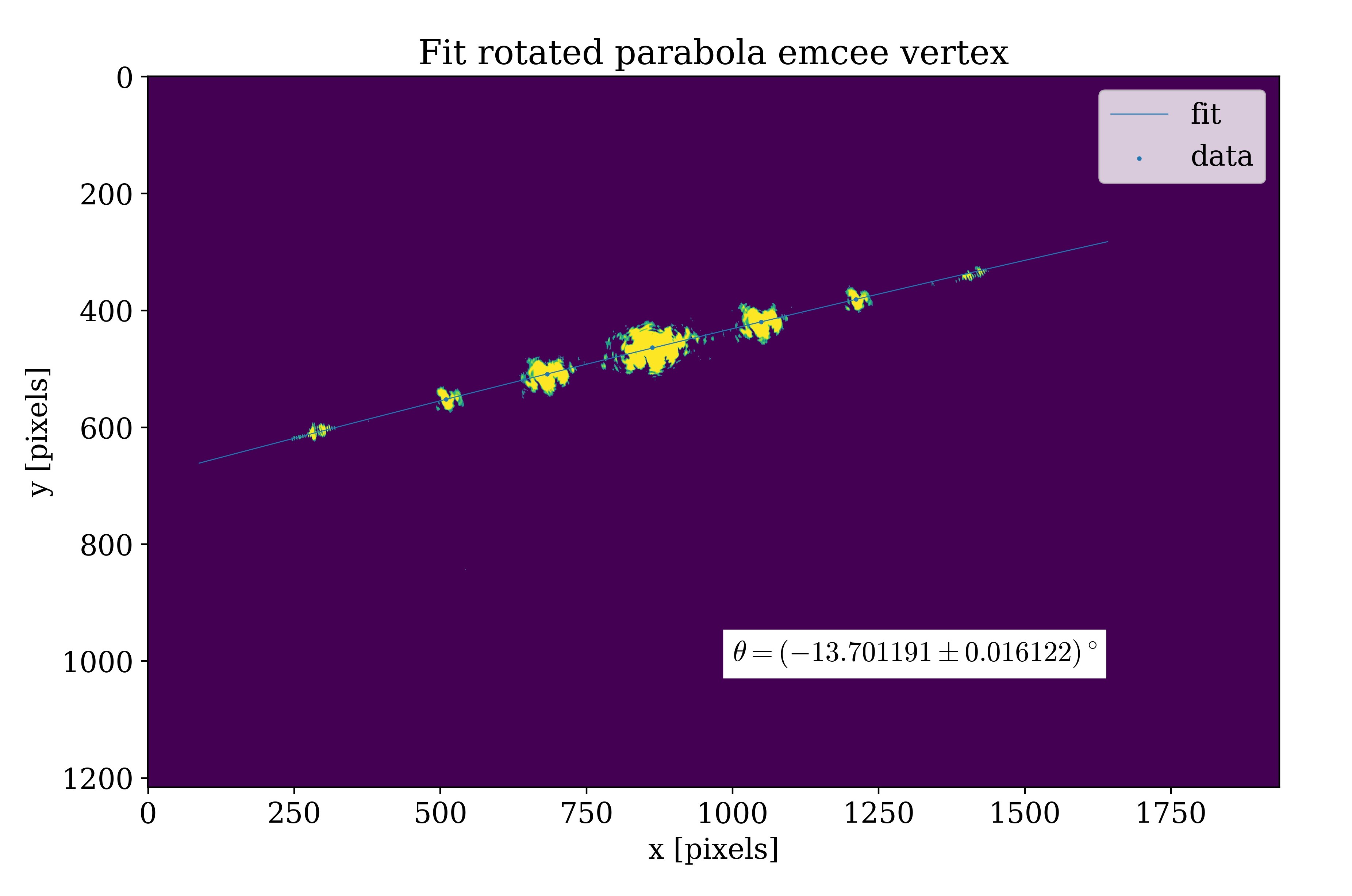}
    \caption{Fit of the diffraction pattern image with the rotated parabola, using the vertex $(x_V,y_V)$ as free parameters. The systematic uncertainty on $\theta$ is smaller than $0.05^\circ$.}
    \label{fig:rot_parabola_fit_vertex}
\end{figure}

\begin{figure}
    \centering
    \includegraphics[width=\columnwidth]{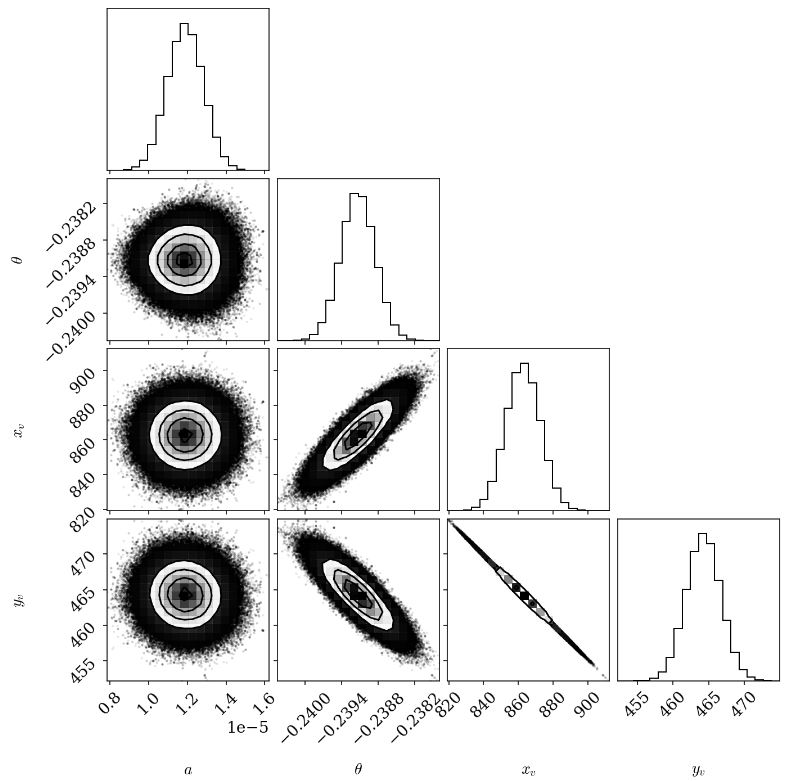}
    \caption{Corner plot of the \textit{emcee} fit using the vertex $(x_V,y_V)$ as free parameters (Fig~\ref{fig:rot_parabola_fit_vertex}). We can notice a correlation between the parameters $\theta$ and $(x_V,y_V)$, since the position of the vertex strongly depends on the orientation of the diffraction pattern.}
    \label{fig:corner_vertex}
\end{figure}

Fig.~\ref{fig:rot_parabola_fit_vertex} and Fig.~\ref{fig:corner_vertex} show (respectively) typical results for the second fitting method and their corner plot. We have a correlation between the angle $\theta$ and $(x_V,y_V)$, as expected, since the orientation of the parabola strongly depends on the position of its vertex.
\\ 
\\

\begin{table}[]
\centering
    \begin{tabular}{|c|c|c|c|}
    \hline
    & $\overline{\sigma}_{syst}$ & 
    ${\sigma}_{stat}$ & $\overline{\sigma}_{\theta,tot}$ \\
    \hline
        $1^{st}$ method & $0.0075^\circ$ & $0.0498^\circ$ & $0.0504^\circ$\\
        $2^{nd}$ method & $0.0134^\circ$ & $0.0595^\circ$ & $0.0610^\circ$\\
    \hline
    \end{tabular}
    \caption{From left to right: average systematic uncertainty, statistical uncertainty and average total uncertainty on the diffraction pattern inclination angle in the camera plane. These uncertainties are computed over $10$ different images with the same orientation of the polarizer. The results of the two fitting methods are compared.}
    \label{tab:sigma results}
\end{table}

\subsubsection{Photogrammetry}
\label{sec:Photogrammetry}

The objective of photogrammetry is to determine the orientation of the CCD camera plane with respect to an absolute coordinate system, in order to get a final independent estimation of the polarization angle that should be measured by the NIKA2 detectors.\newline
To achieve this, the idea is to randomly place some ground references (hereafter landmarks) around the IRAM 30m telescope and to detect them with the refractor telescope mounted in the COSMOCal box (section~\ref{sec:optical design}), similarly to what proposed by \cite{Dunner2020}.\newline
To succeed in photogrammetry, we need accurate GPS measurements of the position of the source and the landmarks, together with the identification of the origin of the coordinate system.
The precisely measured GPS coordinates of the NIKA2 receiver cabin will allow us to identify the origin of our reference frame. At this point, the 3D positions of the landmarks in the ENU coordinate system centered on the NIKA2 receiver cabin can be established. A dedicated software, developed for the \href{https://gabrielecoppi.github.io/projects/protocalc}{PROTOCALC project}, will enable us to make the link between the 3D positions of the landmarks and their 2D positions in the camera plane. The final outcome of the software will be the three Euler angles (yaw, pitch, and roll) describing the orientation of the camera plane with respect to the ENU coordinate system centered on the NIKA2 receiver cabin. In particular, the roll angle is directly related to the polarization orientation, since it will give the orientation of the camera w.r.t. the optical axis. \newline
This procedure will allow us to get an independent measurement of the polarization orientation that is expected to be measured by the NIKA2 detectors. This approach seems to be very promising given the results obtained by \cite{Dunner2020} and \cite{coppi2022}.

\section{Full-system tests with a KIDs based instrument}
\label{sec:lab_tests_grenoble}

To establish a robust validation of the COSMOCal prototype before the observing campaign at the IRAM 30m telescope, in February 2024 we performed a test week with the fully assembled prototype and a similar instrument to NIKA2. The purpose of the tests was to assess the functionality of the assembled COSMOCal prototype (see the right panel of Fig.~\ref{box inside}) and to establish a data analysis framework using kinetic inductance detectors (KIDs) as receivers, which are also employed in the NIKA2 camera.

During this one-week session, we were able to confirm the ability to measure the polarization angle independently and to evaluate these measurements against those derived from the data analysis of the  measurements with the KIDs camera.

\subsection{Experimental setup}

The receiver used for these measurements is derived from the KISS instrument \citep{Fasano2020}. This versatile instrument offers various configurations, serving as a photometer, polarimeter, or Fourier Transform Spectrometer (FTS). It is composed of a dilution refrigeration cryostat housing multiple screens at different temperatures, three focusing lenses, and a pair of KID arrays located at the coldest stage, approximately 150 mK. The operational principle of the KID detectors is based on the concept of resonance frequency. Essentially, each KID works as a RLC circuit, precisely tuned to a specific resonance frequency. When a photon with energy exceeding the working temperature of the KIDs is received, it induces a shift of its resonance frequency. Analyzing this frequency shift allows us to extract the properties of the detected signal.
In terms of electronic signal processing, we can derive two parameters for each KID: the in-phase component (I) and the quadrature component (Q) of the signal. Specifically, for our measurements, we employed Lumped Element Kinetic Inductance Arrays (LEKIDs), identical to those integrated in the NIKA2 cryostat. LEKIDs offer exceptional sensitivity, responsivity, decay constants, and broad-band applications, coupled with a relatively straightforward fabrication process and multiplexing capabilities \citep{Catalano2020}.\newline
At the coldest stage of the cryogenic system, the two KID arrays are positioned—one in a transmission configuration and the other in a reflection configuration—divided by a linear polarizer tilted at a $45^\circ$ angle relative to the optical axis. For laboratory tests conducted, only the transmission array was used.

The 418 pixel KIDs array used has been previously tested showing good performances for laboratory measurements purposes. In Tab.~\ref{tab:perf_kids}, a summary of the performance of the array is reported. 
\\
\\
\begin{table}[]
    \centering
    \begin{tabular}{|c|c|}
    \hline
     Cryogenic run & NICA V10.1 \\
    \hline
    Silicon wafer & 321 $\mu$m  \\
    Central frequency & 150 GHz \\
    Resonances & 87\% \\
   $Q_{tot}$ & 16555 $\pm$ 4232 \\
    Responsivity & (641 $\pm$ 80) Hz/K \\
    Noise & (11 $\pm$ 2) $Hz/\sqrt{Hz}$ \\
    Beam & (6 $\pm$ 1) mm \\
    \hline
    \end{tabular}
    \caption{Performances of the KIDs array used for laboratory tests at the LPSC laboratory.}
    \label{tab:perf_kids}
\end{table}

\subsubsection{Optical chain}
The KIDs array used for the laboratory tests of the COSMOCal mm source is optimally designed for measurements at a wavelength of 1 mm. It is naturally sensitive to the entire spectrum, with the only restriction being the superconducting gap of aluminum at 90 GHz, which sets the lowest observable frequency. The broad absorption range of the array makes it necessary to employ filters to fine-tune the bandwidth according to the source’s emission.
In our current setup, with a source emitting at 265 GHz, we have incorporated an optical low-pass filter (LPF) within the 1~K cryogenic stage to restrict the array’s bandwidth around this frequency. Additionally, a 54~mm pupil was positioned at the 100~mK stage to enhance the focusing of the signal.
The optical system further includes various focusing elements. At the 4~K stage, another LPF at 11 cm$^{-1}$, a high-density polyethylene (HDPE) lens, and a field stop with a diameter of 108~mm have been introduced. Moving up to the 50~K stage, another LPF at 12 cm$^{-1}$ with an anti-reflection coating was added, followed by a thermal filter at the 150~K stage. Finally, an external HDPE window lens was placed at the entrance of the cryostat.

The instrumental setup also includes a set of three linear polarizers lying on the same optical axis, but with three different functions and uses. The first polarizer (P1) located in the COSMOCal box is the crucial element of the optical system, as described in Sec.~\ref{sec:optical design}. P1 can be rotated along the optical axis, at an angle that is roughly estimated by eye. In the complete optical chain of the experimental setup, there is a second polarizer (P2) that is placed at the entrance of the cryostat and can be rotated along the same axis with a precise angle, readable from the graduated grid imprinted on the mechanical support in which it is mounted. Finally, the third and last polarizer (P3) is placed in the coldest cryogenic stage, splitting the signal onto two arrays. Of course, this one cannot be moved during measurements.

\subsubsection{Mechanical setup}

In the COSMOCal box, a mechanical chopper modulates the emission from the millimeter source. This chopper includes an aluminum plate that moves vertically and is adjustable in frequency through dedicated software. A layer of absorbing eccosorb is positioned between the source and the chopper. This layer serves a dual purpose: first, to attenuate the source power and second, to absorb any waves that may be reflected by the chopper back towards the source. This configuration ensures precise modulation of the emission and minimizes unwanted reflections, enhancing accuracy during the measurements.

\begin{figure}
    \centering
    \includegraphics[width=\columnwidth]{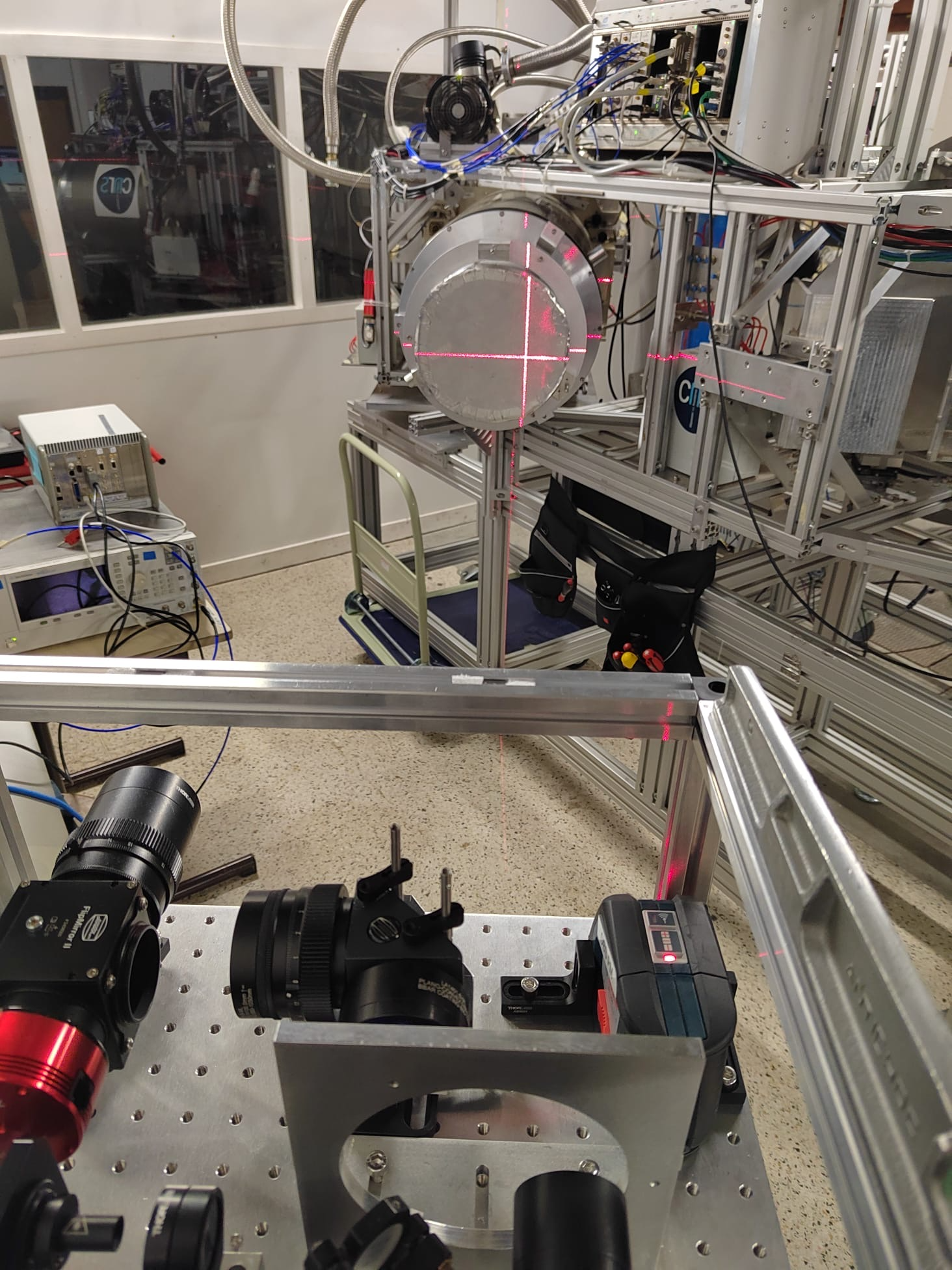}
    \caption{Picture taken during the laboratory measurements at the LPSC laboratory. The cross laser displayed on the cryostat window shows the alignment measurements.}
    \label{fig:crosslaser}
\end{figure}

The entire COSMOCal box was mounted on a tripod and positioned in front of the cryostat entrance, approximately 2 meters away and at a height of 135 cm. An essential aspect of these measurements involved aligning the source w.r.t the cryostat, ensuring accurate data collection. To achieve this alignment, we used a cross-laser, as depicted in Fig.~\ref{fig:crosslaser}. This method facilitates precise alignment, which is crucial for obtaining reliable measurements.

\subsection{Data model and methods}

The starting point for these measurements is a data model designed to simulate the experimental conditions expected in the laboratory. To this end, we used the Stokes and Mueller polarization modeling formalism to compute the theoretical signal expected to be received by the detectors. Here we assume perfect polarizers (complications are addressed in Sec.~\ref{subsec:results_KIDs}).

In this formalism, a perfect polarizer whose transmission axis makes the angle $\alpha$ w.r.t the laboratory reference reads:
\begin{equation}
\begin{array}{rcl}
M^{\prime} &=& M(-2\alpha) \cdot M_{pol} \cdot M(2\alpha) \\
&=& \frac{\displaystyle 1}{\displaystyle 2}
\left[
\begin{array}{ccc}
    1 & \cos 2\alpha & \sin 2\alpha \\
    \cos 2\alpha & \cos^2 2\alpha & \cos 2\alpha \sin 2\alpha \\
    \sin 2\alpha & \cos 2\alpha \sin 2\alpha & \sin^2 2\alpha
\end{array}
\right]
\end{array}
\label{eq:mueller}
\end{equation}

Where $M_{pol}$ is the matrix of an ideal polariser.
In our model, we consider Eq.~\ref{eq:mueller} and replace $\alpha$ with: i) $\psi$ to simulate the COSMOCal's polarizer (P1), ii) $\gamma$ to account for the polarizer (P2) at the entrance of the cryostat and $\beta$ that refers to the cold splitting wire grid (P3). The final signal reaching the detectors is therefore the result of an input Stokes vector S$_{\rm in}$ multiplied by three Mueller matrices: S$_{\rm model}$ = P$_3$($\beta$)$\cdot$ P$_2$($\gamma$)$\cdot$ P$_1$($\psi$)$\cdot$ S$_{\rm in}$. Considering only the first row in P$_3$($\beta$), that gives the amplitude of the signal detected, we obtain:
\begin{equation}
\begin{array}{rl}
S_{\text{model}} = & 1 + \frac{1}{2} \cos 2(\beta - \psi) + \cos 2\gamma (\cos 2\beta + \cos 2\psi) \\
& + \sin 2\gamma (\sin 2\beta + \sin 2\psi) + \frac{1}{2} \cos 4\gamma \cos 2(\beta+\psi) \\
& + \frac{1}{2} \sin 4\gamma \sin 2(\beta+\psi)
\end{array}
\label{eq:sig_model}
\end{equation}

\begin{figure}
    \centering
    \includegraphics[width=\columnwidth]{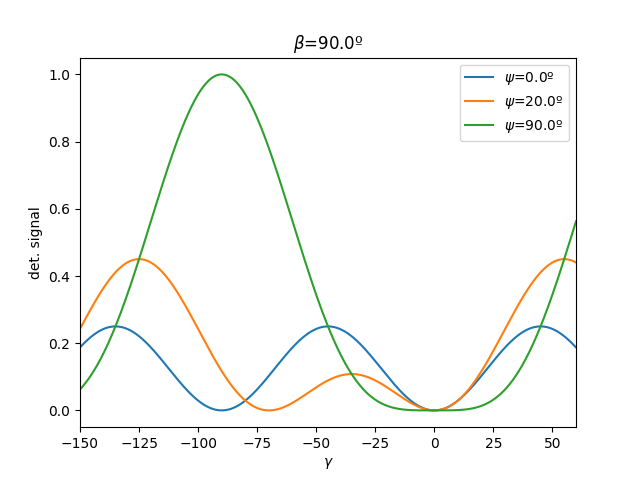}
    \caption{Model signal derived through Mueller formalism for three different configurations of the COSMOCal's polarizer (P1), as a function of $\gamma$, corresponding to the rotating polarizer (P2).}
  \label{fig:mod_pol}
\end{figure}

Due to the mechanical design of the cryostat, $\beta$ is fixed at $90^\circ$, while $\gamma$ varies between [0,$\pi$]. $\psi$ is fixed for each set of measurements. A sample of three different examples is shown in Fig.~\ref{fig:mod_pol}.

\subsubsection{Data acquisition strategy}

Based on the model outlined in the previous section, we have performed three sets of measurements, each corresponding to a different angle of the polarizer P1 $\psi \simeq$ [$0^\circ, 45^\circ, 70^\circ$] w.r.t. the vertical position of the mark on P1 indicating the output direction of the polarization, approximately. The exact values are derived {\it a posteriori} through an accurate analysis of the diffration pattern, see Sec.~\ref{sec:Diffraction pattern analysis}. Next, for each fixed position of $\psi$, we rotated P2 through a range of 10 different angles, spaced at intervals of $20^\circ$, spanning -130$^\circ$ to +50$^\circ$. This polarizer P2 acts as an analyzer and its angle is known with a mechanical precision of better than 1$^\circ$ due to the marks graduations on its mechanical mount.\newline

During laboratory tests, we noticed that the rotation of the COSMOCal input polarizer (P1) produced significant changes in the response of the KIDs, due to the different amplitude of the polarization signal. Consequently, a re-tuning procedure was needed to get them back to their working resonance frequency after any P1 rotation angle change. Finally, a mechanical chopper modulated the signal at 0.16~Hz before P2, enabling the lock-in detection of the signal and rejecting low frequency electronic noise.

\subsection{COSMOCal system's independent results}

As reference on the precise knowledge of the COSMOCal output, the polarization angle was independently measured through the diffraction pattern analysis method described in Sec.~\ref{sec:Diffraction pattern analysis}. Due to the laboratory setting, photogrammetry was not available. Consequently, we employed a plumb line to ascertain the camera roll angle, meticulously aligning it with a suitable reference. However, this approach emerged as our main source of uncertainty.
\newline
Through the diffraction pattern analysis, we obtained the following results for the three rotating angles of P1: $\psi_{1,DP}=(351.295 \pm 0.023)^\circ$, $\psi_{2,DP} = (23.477 \pm 0.041)^\circ$ and $\psi_{3,DP} = (51.662 \pm 0.011)^\circ$. For these angles, we need to apply the correction for the roll angle, estimated from the reference lines aligned with the plumb line. \newline
The reference lines have been identified with an OpenCV algorithm from the pictures of a millimeter paper aligned with the plumb line, acquired with COSMOCal CCD camera and an apposite lens. The roll angle was calculated as the median of the inclination distribution of the detected lines and the roll angle uncertainty was estimated through $\sigma_G = 0.7413 \times (q_{75} - q_{25})$, where $q_{75}$ and $q_{25}$ are, respectively, the $75^{th}$ and $25^{th}$ percentiles of the distribution. We used the median and $\sigma_G$ as estimators since they are more robust than the mean and standard deviation, being less affected by outliers, which can arise from the line detection algorithm mentioned above. This results in the following roll angle assessment:
\begin{equation}
    \varphi_{roll} = (13.062 \pm 0.065)^\circ
\end{equation}
Applying this offset to the angles resulting from the diffraction pattern analysis, we obtained the polarization angles reported in Tab.~\ref{tab:res_fit_pure}. As mentioned above, the largest source of uncertainty in the polarization angle measurements arises from the estimation of the roll angle. Indeed, although the results shown in Tab.~\ref{tab:res_fit_pure} exhibit uncertainty about the polarization angle below the requirement of $\Delta\psi < 0.1^{\circ}$, this is a pure data analysis outcome. Actually, our result is strongly affected by the precision in the alignment of the plumb line with the millimeter paper used for line detection, which could not be tightly constrained in the laboratory setup. The results presented here show our capability to obtain an independent knowledge of the polarization angle output from the COSMOCal prototype. This will certainly be better determined during the IRAM 30m test campaign using photogrammetry.  

%%%%%%%%%%%%%%%%%%%%%%%%%%%%%%%%% Results %%%%%%%%%%%%%%%%%%%%%%%%%%%%%%%%%%%%%%%%%%%%%%%%%%

\subsection{Results using KIDs data}
\label{subsec:results_KIDs}

At the end of the measurement campaign, we have acquired a comprehensive data set, which we analyzed in several stages. Initially, we turned the (I,Q) raw data into a phase timeline that is proportional to the optical total power that reaches the detectors. Once this streamlined data set was derived, we performed a more complex noise treatment and data reduction process to obtain refined data points suitable for thorough analysis and fitting to the model function.

\subsubsection{Raw data treatment}

First of all, raw data are expressed in binary format and we cannot directly treat data in (I,Q), as they are. So, we convert the (I,Q) data into phase data applying Eq.~\ref{eq:IQ}.

\begin{equation}
    \phi = \arctan \frac{Q}{I}
    \label{eq:IQ}
\end{equation}

\begin{figure}
    \centering
    \includegraphics[width=\columnwidth]{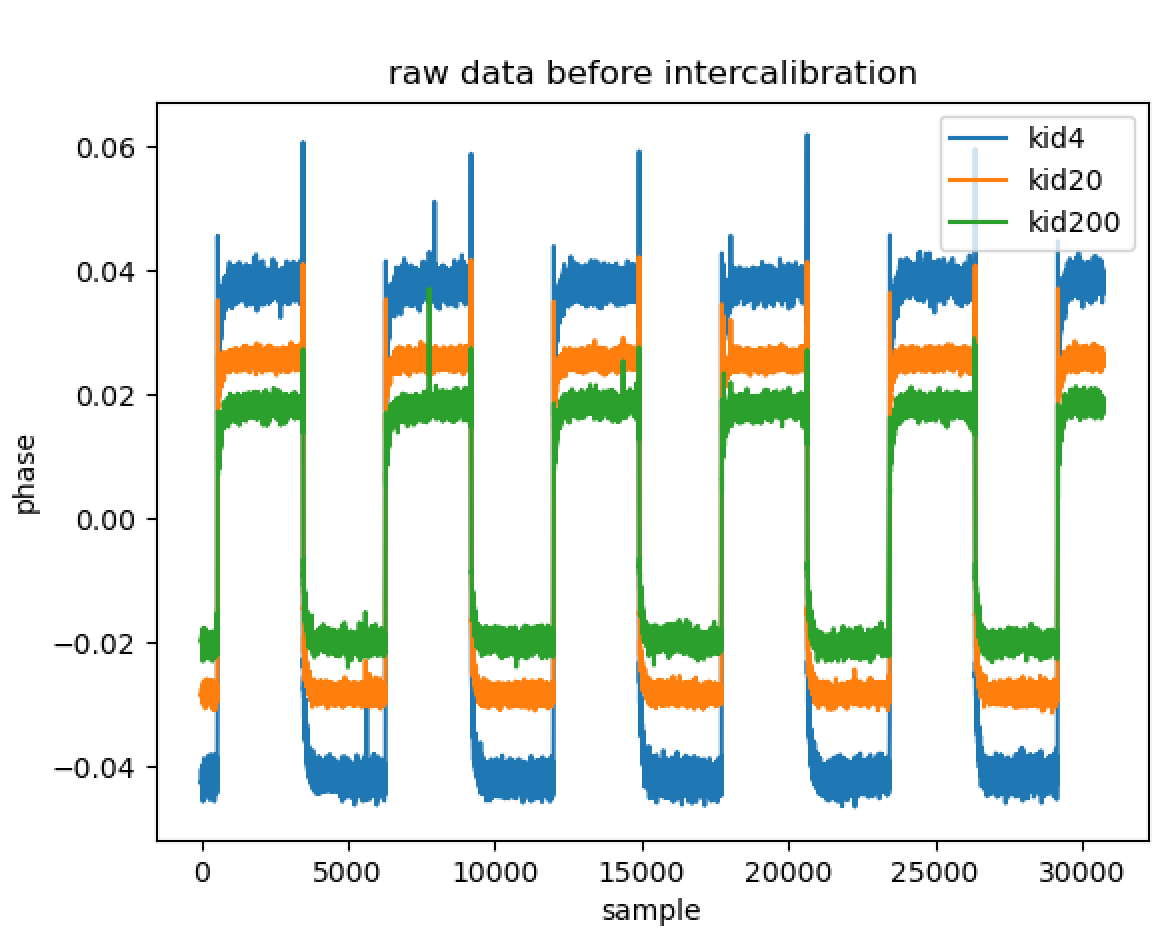}
    \caption{Raw phase data for a set of three KIDs (4, 20, 200) before inter-calibration.}
    \label{fig:raw_phase}
\end{figure}

\begin{figure}
    \centering
    \includegraphics[width=\columnwidth]{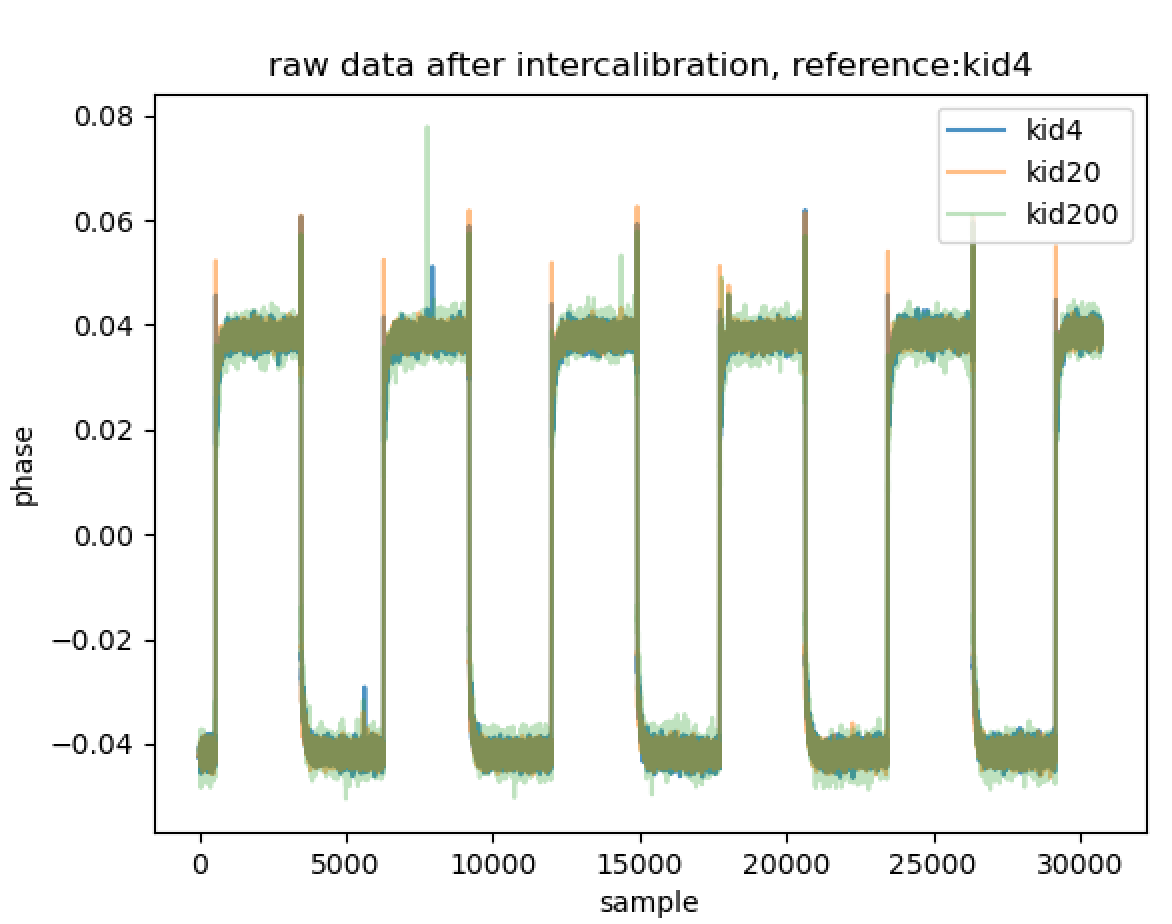}
    \caption{Raw phase data for a set of three KIDs (4, 20, 200) after inter-calibration.}
    \label{fig:raw_intercalib}
\end{figure}

Once we obtained the data of the phase, we proceeded to clean them from glitches and disturbances caused by changes in the signal caused by the chopper. We did so by subtracting the median of the signal for each plateau, corresponding to ups and downs of the chopper. Doing so for each KID, we account for the fact that KIDs are not equally responsive and, hence, they do not record the same signal amplitude. Therefore, an intercalibration of the KIDs was necessary, finally considering KID n.4 as a reference. The difference between raw and intercalibrated data is shown in Figs.~\ref{fig:raw_phase} and \ref{fig:raw_intercalib}.

We computed the average signal for all the up and down blocks from the raw data and derived a single data point for each time series. This process provided ten data points per KID for each angle of $\psi$. \newline
The following step involved fitting these ten data points with the model function, incorporating minor corrections. The first correction accounts for the tilted position of P2 w.r.t. to its optical axis of $\sim$ 10$^\circ$, which is enough to prevent or at least significantly reduce reflections. Due to this tilt, the transmitted polarization is the component projected onto the plane perpendicular to the optical axis. We then calculated the corrected angle to be used in the model function in replacement of $\cos \gamma$, as shown in Eq.~\ref{eq:corr_tilt}, where $\tau$ is the tilt angle.

\begin{equation}
    \cos \gamma_{tilt} = \frac{\cos\gamma \cdot \cos\tau}{\sqrt{\cos\gamma ^2 \cdot \cos\tau^2 + \sin\gamma^2}} 
\label{eq:corr_tilt}
\end{equation}

The second correction that we should consider is due to the fact that all polarizers used are not actually ideal, which means that they do not transmit one pure component along the opposite direction of the wires, but some other minor components of the signal are transmitted as well. To apply this correction, the matrix in Eq.~\ref{eq:mueller} should be modified, adding the appropriate constants. The corrected version of the Mueller matrix assuming a non-ideal polarizer is shown in Eq.~\ref{eq:Mueller_not_ideal}.
\[
\small
M= \left[
   \begin{array}{ccc}
    K & k\cos 2\alpha & k\sin 2\alpha  \\
    k\cos 2\alpha & K\cos^2 2\alpha + q\sin^2 2\alpha & (K-q)\cos2\alpha \sin2\alpha  \\
    k\sin 2\alpha & (K-q)\cos 2\alpha \sin 2\alpha & K\sin^2 2\alpha + q\cos^2 2\alpha  \\
\end{array}
\right]
\]
\label{eq:Mueller_not_ideal}

However, by applying the non-ideal polarizer corrected Mueller matrix we realized that it only accounts for minor corrections with respect to the original signal, while adding non-negligible complexity to the fitting function, due to multiple constants to be handled. Therefore, we opted for a simpler model, assuming all ideal polarizers.

%\np{Did we actually do this ? If yes, what are these constants and how were they determined ?} \newline
Once the tilt angle correction reported in Eq.~\ref{eq:corr_tilt} was included in the model, we could fit the data with the final model function, also adding an overall amplitude in front of it (Eq.~\ref{eq:fit_pure}) to account for normalization.

%Once these corrections have been included in the model, we can eventually fit the data with the model function including the discussed corrections and an overall amplitude, therefore using an equation of the form of Eq.~\ref{eq:fit_pure}, where $S_{corr}$ takes into account the two corrections. %\np{this needs to be rephrased : it loops on mentionning "the" corrections, but they have not be specified.}

\begin{equation}
    S_{tot}(\gamma_{tilt}, \psi) = A \cdot S_{corr}(\gamma_{tilt},\psi)
\label{eq:fit_pure}    
\end{equation}

\subsubsection{Polarization angle fit}

\begin{figure}
    \centering
    \includegraphics[width=\columnwidth]{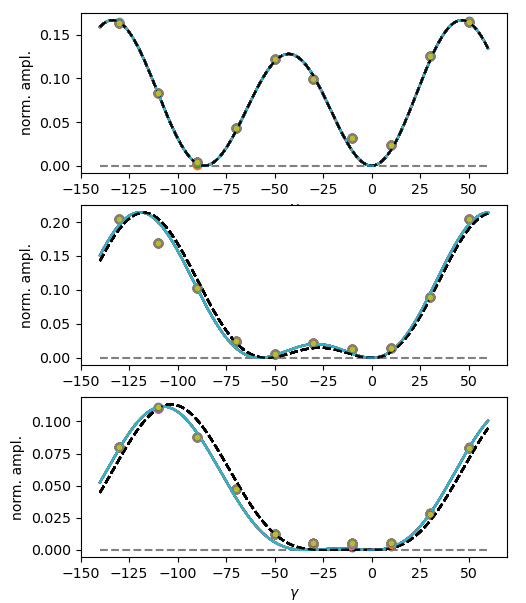}
    \caption{For all plots, the green dots actually hide the 19 other best KIDs that were selected and whose data are superposed. The dashed black line is the fit of the data obtained by imposing the $\psi$ angles derived by the diffraction pattern analysis. The blue line is again a superposition of the 20 best fit of the KIDs of the data, this time leaving $\psi$ as a free parameter. The three panels show data for the three sets of measurements, corresponding to $\psi$ = [0$^{\circ}$, 45$^{\circ}$, 70$^{\circ}$].}
    \label{fig:fit_noparam}
\end{figure}

The initial results obtained from fitting the data for all KIDs varied in quality across the array composed of 317 working detectors, each one characterized by a specific resonance frequency and location with respect to the center of the array. The variation of quality among different pixels can be due to different reasons: for example, positional effects can affect the performances, since the detectors located at the edge of the array may experience different environmental conditions compared to the ones placed in the array's center. In addition, there could be variations in noise levels and cross-talk effects between adjacent detectors that impact the quality of their response. Finally, some of them might be structurally less performing and precise than others, due to variations that can occur during the fabrication process. For these reasons, in particular in such non-optimal laboratory test conditions, it is common and good practice to extract a selection of the best pixels in order to obtain the best results.
%\np{It would be good to provide a clue on why, otherwise the reader might get the idea that KIDs are not good}. 
Consequently, we evaluated the goodness of fit by calculating the values of $\chi^2$, ultimately selecting the top 20 KIDs with the lowest $\chi^2$. We then performed a second round of fitting exclusively for these 20 KIDs, resulting in an initial estimate of the two fitted parameters: the amplitude and the angle $\psi$. The data points and fitted curves are shown in Fig.~\ref{fig:fit_noparam}.

The fitted estimates of A and $\psi$ are provided in Tab.~\ref{tab:res_fit_pure}.

\begin{table}[]
    \centering
    \begin{tabular}{|c|c|c|c|}
    \hline
     & A & $\psi$ & $\psi_{DP}$\\
    \hline
    set1 & 0.146 $\pm$ 0.002 & (3.8 $\pm$ 0.3)$^\circ$ & (4.357 $\pm$ 0.069)$^\circ$\\
    set2 & 0.091 $\pm$ 0.002 & (31.9 $\pm$ 0.9)$^\circ$ & (36.539 $\pm$ 0.077)$^\circ$\\
    set3 & 0.0334 $\pm$ 0.0005 & (55.9 $\pm$ 0.9)$^\circ$ & (64.724 $\pm$ 0.066)$^\circ$\\
    \hline
    \end{tabular}
    \caption{Fit results for amplitude A and polarization angle $\psi$ for the three sets of measurements, compared to the values of $\psi_{DP}$ obtained from diffraction pattern analysis.}
    \label{tab:res_fit_pure}
\end{table}

We can directly compare these results with the angles derived through the analysis of the diffraction pattern, which are listed in Tab.~\ref{tab:res_fit_pure}. Clearly, there is a notable difference among these values; in fact, the percentage error is around 15\% for all of them, while the absolute error is as high as 9$^\circ$ for the third set of measurements. 

The experimental setup, combined with the source characteristics, presents some constraints that can generate disturbing effects. In particular, the power of the source is designed to suit a telescope configuration, notably for the IRAM 30m tests. This power exceeds the power tolerated by KIDs detectors in the laboratory configuration due to the near field. Therefore, it was expected that parasitic reflections would show up. The most probable source of reflection is coming from the COSMOCal polarizer P1. The fraction of the source radiation that is not transmitted and directly sent to the cryostat is reflected and may bounce back in the room up to finally entering into the cryostat. This component has a 90 $^\circ$ phase compared to the transmitted component. We account for this additional component in our fitting model. Because this latter also passes through the entrance analyzer and the splitter, its form is the same as that in Eq.~\ref{eq:sig_model}. Our fitting model thus reads:

%One possible way to correct this discrepancy is adding a parasitic component to the model, which takes into account the possible reflections of the signal on the mechanical chopper and/or in the laboratory room. The modeled parasitic component is added to the pure signal, having the same analytic form, so that the final fitting function would be the one shown in Eq.~\ref{eq:parasitic}, where the angle $\Bar{\psi} = \psi + 90^\circ + \phi$. 
%In practice, we assume that the parasitic component is modulated in phase opposition to the pure component. This assumption results in a 90$^\circ$ phase shift, which is incorporated into the determination of the new angle.

\begin{equation}
    S_{tot} = A \cdot S_{pure}(\gamma, \psi) + B \cdot S_{para}(\gamma, \bar{\psi})
\label{eq:parasitic}
\end{equation}

\begin{figure}
    \centering
    \includegraphics[width=\columnwidth]{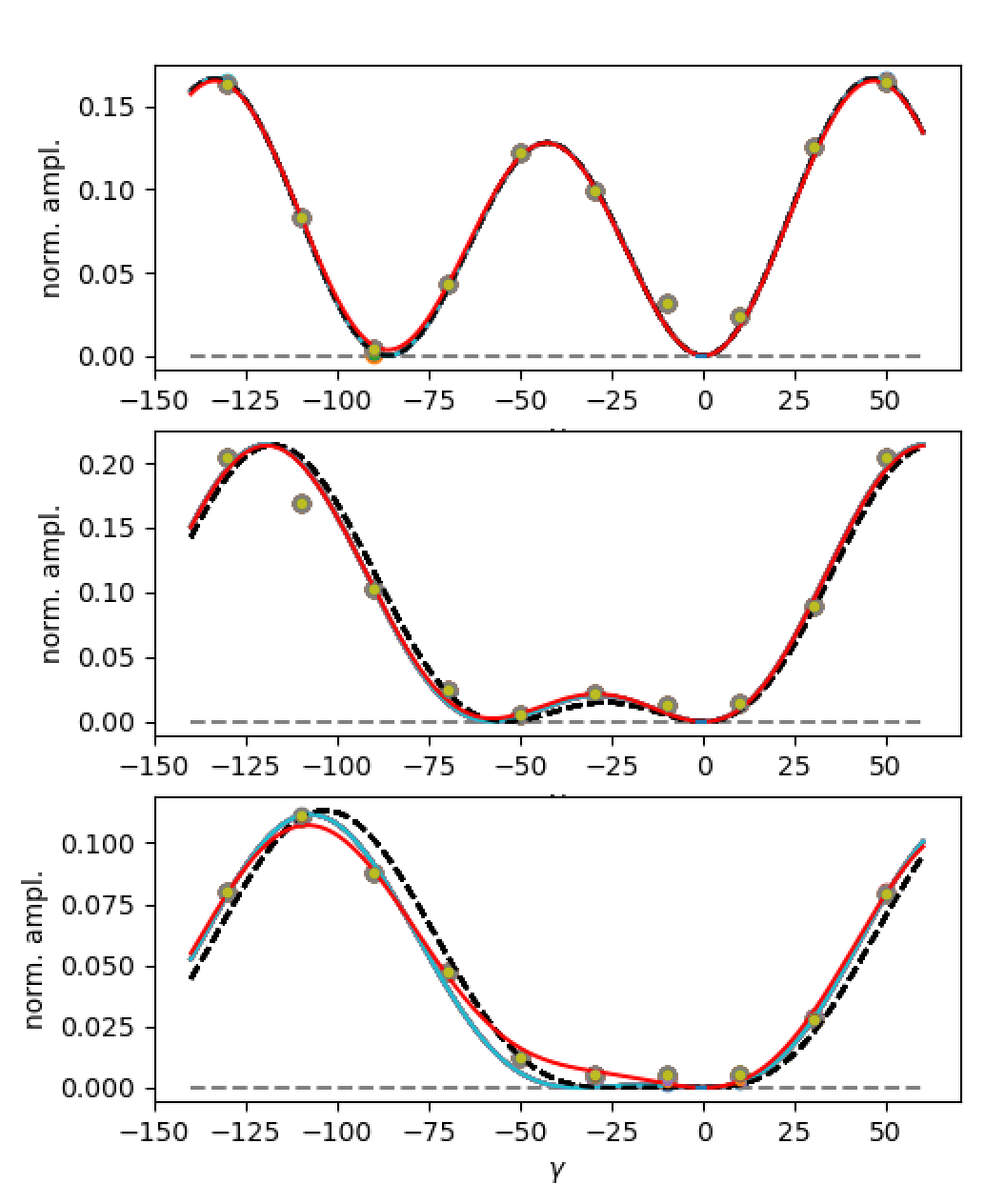}
%    \caption{For all plots, the green dots represent the superposition of the 20 best KIDs data. The dashed black line is the fit of the data obtained by imposing the $\psi$ angles derived through the diffraction pattern analysis. The blue line is again a superposition of the 20 best KIDs fit of the data, this time leaving $\psi$ as a free parameter. The red line is the fit calculated including the parasitic component. The three panels show data for the three sets of measurements, corresponding to the three $\psi$ angles.}
\caption{Same curves as Fig.~\ref{fig:fit_noparam}, with the added fit including the parasitic component in red.}
    \label{fig:fit_paras}
\end{figure}

Finally, we again fit the data with this new parasitic component, and we estimate the values for the four fitted parameters. The plot of the new fit, accounting for parasitic components, is shown in Fig.~\ref{fig:fit_paras}. The new fitted parameters are listed in Tab.~\ref{tab:res_fit_paras}. This fit was performed by running an $emcee$ MCMC in order to eliminate possible degeneracies among the fit parameters. We set the following priors for the 3 different polarization angles ($\psi$): [3,6], [30,40], [55,65] degrees, while for the phase angle ($\phi$) the priors are set to [50,90] degrees for all 3 cases. According to the results obtained, we have concluded that we are not limited by the priors and that this choice of priors is justified by the previous knowledge of the fitted polarization angles acquired through the fit without parasitic components.

\begin{table*}[t] % [t] for top of the page
    \centering
    \begin{tabular}{|c|c|c|c|c|c|}
        \hline
    $\psi_{\text{box}}$ & $\psi_{det}$ & A & B & $\phi$ \\
    \hline
    (4.357 $\pm$ 0.069)$^\circ$ & (4.3 $\pm$ 0.7)$^\circ$ & 0.146 $\pm$ 0.002 & 0.0116 $\pm$ 0.0001 & (77 $\pm$ 5)$^\circ$  \\
    (36.539 $\pm$ 0.077)$^\circ$ & (35.6 $\pm$ 0.9)$^\circ$ & 0.069 $\pm$ 0.003 & 0.023 $\pm$ 0.005 & (78 $\pm$ 6)$^\circ$\\
    (64.724 $\pm$ 0.066)$^\circ$ & (63.8 $\pm$ 0.8)$^\circ$ & 0.027 $\pm$ 0.003 & 0.012 $\pm$ 0.004 & (76 $\pm$ 5)$^\circ$\\
    \hline
    \end{tabular}
    \caption{First two columns show the comparison between the COSMOCal independent measurements of the polarization angle ($\psi_{box}$) and the ones measured through analysis of KIDs data ($\psi_{det}$). Third, fourth and fifth columns show the results of the MCMC fit.}
    \label{tab:res_fit_paras}
\end{table*}

\subsubsection{Final remarks}

The final results for the fitted parameters are still not completely satisfactory, although we notice that the addition of a parasitic component plays a significant role and reduces discrepancies between the values of the polarization angle $\psi$ estimated with different methods. However, we still lack a complete understanding of the physical nature of these parasitic signals that contribute to the pure signal of the source. We emphasize that these are due to the environment in which we have performed these measurements that is not designed for accurate millimeter optics characterization.

\section{Conclusions and perspectives}
\label{sec:conclusions}

In response to the demanding requirements for precision in astrophysics and cosmology research, the COSMOCal project is dedicated to establishing an unparalleled standard for calibrating polarization angles in microwave experiments operating within the critical $\sim$ 90-300 GHz range. Although the primary scientific motivation behind this technological endeavor is to enable CMB experiments to accurately detect primordial \bmodes, the project also has the potential to expand its scientific scope. For example, by constraining physical phenomena beyond the standard cosmological model, such as cosmic birefringence.

The article details the development of the COSMOCal prototype, designed to match the sensitivity and detection strategy of the NIKA2 camera installed at the IRAM 30m telescope. The prototype includes a millimeter source consisting of a chain of radio-frequency (RF) components, while the optical system consists of various optical elements necessary to obtain an independent measurement of the polarization angle during observations, see Fig.~\ref{box inside}. Preliminary laboratory measurements, performed at the LPENS and {\it Observatoire de Paris}, yielded promising results, demonstrating that the system can achieves a polarization angle precision of 0.06$^\circ$.

Further laboratory tests performed at the LPSC institute in February 2024, with the fully assembled COSMOCal prototype interfaced with a Kinetic Inductance Detectors-based camera confirmed these results but revealed technical challenges due to the KIDs detection of a stray microwave signal and uncertainties in estimating the COSMOCal camera roll angle. Nevertheless, through rigorous data analysis, we have been able to gain valuable insight from our measurements. In particular, independent measurements of the roll angle of the COSMOCal polarizer and the signal polarization with the KIDs camera agree within 2-3\% in absolute value. This difference is probably due to the limitations of the laboratory setup, which does not provide optimal cancellation of reflections. Furthermore, the uncertainties in the camera measurements dominate the error budget. 

Future tests at the IRAM 30m telescope should overcome laboratory technical constraints providing a more reliable setup for millimeter wavelengths measurements. However systematic effects can arise from the interface between the COSMOCal system and the entire optical chain of the telescope. The overarching aim of this forthcoming campaign is to ensure the fidelity of data gathered by the NIKA2 camera regarding the orientation of the COSMOCal polarization signal across the entirety of the telescope's optical chain. Although this initial assessment is pivotal for validating our assumptions and gauging the present calibration of the NIKA2 polarization system, the comprehensive exploration and mitigation of all systematic effects, including instrumental polarization and optical efficiency, necessitate having the COSMOCal system in space, in the far field for such large telescopes. 
%Such deployment would optimize the system performance and accuracy, enabling as well enhanced observations and measurements of astrophysical phenomena with large aperture telescopes.

To address this goal, we have started to work on the design of a space prototype comprising a microwave source emitting a polarization signal at three frequencies within the 90-300 GHz frequency range. This prototype integrates an optical system optimized for maximizing power throughput and directing light from space towards telescopes in southern Europe and Chile. By exploring the possibility of attaching our system to a SAT COM satellite, we are pioneering a novel collaboration model between private and public entities. This collaboration not only reduces the financial cost of the space mission, but also minimizes the proliferation of spacecraft in orbit around the Earth. 
\section{Acknowledgments}
We acknowledge financial support from CENSUS, Observatoire de Paris-PSL and CNES space agency. A.R. acknowledges financial support from the Italian Ministry of
University and Research - Project Proposal CIR01\_00010.

\bibliography{main}

\begin{thebibliography}{}
\expandafter\ifx\csname natexlab\endcsname\relax\def\natexlab#1{#1}\fi
\providecommand{\url}[1]{\href{#1}{#1}}
\providecommand{\dodoi}[1]{doi:~\href{http://doi.org/#1}{\nolinkurl{#1}}}
\providecommand{\doeprint}[1]{\href{http://ascl.net/#1}{\nolinkurl{http://ascl.net/#1}}}
\providecommand{\doarXiv}[1]{\href{https://arxiv.org/abs/#1}{\nolinkurl{https://arxiv.org/abs/#1}}}

\bibitem[{{Abazajian} {et~al.}(2016){Abazajian}, {Adshead}, {Ahmed}, {Allen}, {Alonso}, {Arnold}, {Baccigalupi}, {Bartlett}, {Battaglia}, {Benson}, {Bischoff}, {Borrill}, {Buza}, {Calabrese}, {Caldwell}, {Carlstrom}, {Chang}, {Crawford}, {Cyr-Racine}, {De Bernardis}, {de Haan}, {di Serego Alighieri}, {Dunkley}, {Dvorkin}, {Errard}, {Fabbian}, {Feeney}, {Ferraro}, {Filippini}, {Flauger}, {Fuller}, {Gluscevic}, {Green}, {Grin}, {Grohs}, {Henning}, {Hill}, {Hlozek}, {Holder}, {Holzapfel}, {Hu}, {Huffenberger}, {Keskitalo}, {Knox}, {Kosowsky}, {Kovac}, {Kovetz}, {Kuo}, {Kusaka}, {Le Jeune}, {Lee}, {Lilley}, {Loverde}, {Madhavacheril}, {Mantz}, {Marsh}, {McMahon}, {Meerburg}, {Meyers}, {Miller}, {Munoz}, {Nguyen}, {Niemack}, {Peloso}, {Peloton}, {Pogosian}, {Pryke}, {Raveri}, {Reichardt}, {Rocha}, {Rotti}, {Schaan}, {Schmittfull}, {Scott}, {Sehgal}, {Shandera}, {Sherwin}, {Smith}, {Sorbo}, {Starkman}, {Story}, {van Engelen}, {Vieira}, {Watson}, {Whitehorn}, \& {Kimmy Wu}}]{2016arXiv161002743A}
{Abazajian}, K.~N., {Adshead}, P., {Ahmed}, Z., {et~al.} 2016, arXiv e-prints, arXiv:1610.02743, \dodoi{10.48550/arXiv.1610.02743}

\bibitem[{{Adam} {et~al.}(2018){Adam}, {Adane}, {Ade}, {Andr{\'e}}, {Andrianasolo}, {Aussel}, {Beelen}, {Beno{\^\i}t}, {Bideaud}, {Billot}, {Bourrion}, {Bracco}, {Calvo}, {Catalano}, {Coiffard}, {Comis}, {De Petris}, {D{\'e}sert}, {Doyle}, {Driessen}, {Evans}, {Goupy}, {Kramer}, {Lagache}, {Leclercq}, {Leggeri}, {Lestrade}, {Mac{\'\i}as-P{\'e}rez}, {Mauskopf}, {Mayet}, {Maury}, {Monfardini}, {Navarro}, {Pascale}, {Perotto}, {Pisano}, {Ponthieu}, {Rev{\'e}ret}, {Rigby}, {Ritacco}, {Romero}, {Roussel}, {Ruppin}, {Schuster}, {Sievers}, {Triqueneaux}, {Tucker}, \& {Zylka}}]{2018A&A...609A.115A}
{Adam}, R., {Adane}, A., {Ade}, P.~A.~R., {et~al.} 2018, \aap, 609, A115, \dodoi{10.1051/0004-6361/201731503}

\bibitem[{{Ade} {et~al.}(2019{\natexlab{a}}){Ade}, {Aguirre}, {Ahmed}, {Aiola}, {Ali}, {Alonso}, {Alvarez}, {Arnold}, {Ashton}, {Austermann}, {Awan}, {Baccigalupi}, {Baildon}, {Barron}, {Battaglia}, {Battye}, {Baxter}, {Bazarko}, {Beall}, {Bean}, {Beck}, {Beckman}, {Beringue}, {Bianchini}, {Boada}, {Boettger}, {Bond}, {Borrill}, {Brown}, {Bruno}, {Bryan}, {Calabrese}, {Calafut}, {Calisse}, {Carron}, {Challinor}, {Chesmore}, {Chinone}, {Chluba}, {Cho}, {Choi}, {Coppi}, {Cothard}, {Coughlin}, {Crichton}, {Crowley}, {Crowley}, {Cukierman}, {D'Ewart}, {D{\"u}nner}, {de Haan}, {Devlin}, {Dicker}, {Didier}, {Dobbs}, {Dober}, {Duell}, {Duff}, {Duivenvoorden}, {Dunkley}, {Dusatko}, {Errard}, {Fabbian}, {Feeney}, {Ferraro}, {Flux{\`a}}, {Freese}, {Frisch}, {Frolov}, {Fuller}, {Fuzia}, {Galitzki}, {Gallardo}, {Tomas Galvez Ghersi}, {Gao}, {Gawiser}, {Gerbino}, {Gluscevic}, {Goeckner-Wald}, {Golec}, {Gordon}, {Gralla}, {Green}, {Grigorian}, {Groh}, {Groppi}, {Guan}, {Gudmundsson}, {Han}, {Hargrave}, {Hasegawa},
  {Hasselfield}, {Hattori}, {Haynes}, {Hazumi}, {He}, {Healy}, {Henderson}, {Hervias-Caimapo}, {Hill}, {Hill}, {Hilton}, {Hilton}, {Hincks}, {Hinshaw}, {Hlo{\v{z}}ek}, {Ho}, {Ho}, {Howe}, {Huang}, {Hubmayr}, {Huffenberger}, {Hughes}, {Ijjas}, {Ikape}, {Irwin}, {Jaffe}, {Jain}, {Jeong}, {Kaneko}, {Karpel}, {Katayama}, {Keating}, {Kernasovskiy}, {Keskitalo}, {Kisner}, {Kiuchi}, {Klein}, {Knowles}, {Koopman}, {Kosowsky}, {Krachmalnicoff}, {Kuenstner}, {Kuo}, {Kusaka}, {Lashner}, {Lee}, {Lee}, {Leon}, {Leung}, {Lewis}, {Li}, {Li}, {Limon}, {Linder}, {Lopez-Caraballo}, {Louis}, {Lowry}, {Lungu}, {Madhavacheril}, {Mak}, {Maldonado}, {Mani}, {Mates}, {Matsuda}, {Maurin}, {Mauskopf}, {May}, {McCallum}, {McKenney}, {McMahon}, {Meerburg}, {Meyers}, {Miller}, {Mirmelstein}, {Moodley}, {Munchmeyer}, {Munson}, {Naess}, {Nati}, {Navaroli}, {Newburgh}, {Nguyen}, {Niemack}, {Nishino}, {Orlowski-Scherer}, {Page}, {Partridge}, {Peloton}, {Perrotta}, {Piccirillo}, {Pisano}, {Poletti}, {Puddu}, {Puglisi}, {Raum}, {Reichardt},
  {Remazeilles}, {Rephaeli}, {Riechers}, {Rojas}, {Roy}, {Sadeh}, {Sakurai}, {Salatino}, {Sathyanarayana Rao}, {Schaan}, {Schmittfull}, {Sehgal}, {Seibert}, {Seljak}, {Sherwin}, {Shimon}, {Sierra}, {Sievers}, {Sikhosana}, {Silva-Feaver}, {Simon}, {Sinclair}, {Siritanasak}, {Smith}, {Smith}, {Spergel}, {Staggs}, {Stein}, {Stevens}, {Stompor}, {Suzuki}, {Tajima}, {Takakura}, {Teply}, {Thomas}, {Thorne}, {Thornton}, {Trac}, {Tsai}, {Tucker}, {Ullom}, {Vagnozzi}, {van Engelen}, {Van Lanen}, {Van Winkle}, {Vavagiakis}, {Verg{\`e}s}, {Vissers}, {Wagoner}, {Walker}, {Ward}, {Westbrook}, {Whitehorn}, {Williams}, {Williams}, {Wollack}, {Xu}, {Yu}, {Yu}, {Zago}, {Zhang}, {Zhu}, \& {Simons Observatory Collaboration}}]{Simons19}
{Ade}, P., {Aguirre}, J., {Ahmed}, Z., {et~al.} 2019{\natexlab{a}}, \jcap, 2019, 056, \dodoi{10.1088/1475-7516/2019/02/056}

\bibitem[{{Ade} {et~al.}(2019{\natexlab{b}}){Ade}, {Aguirre}, {Ahmed}, {Aiola}, {Ali}, {Alonso}, {Alvarez}, {Arnold}, {Ashton}, {Austermann}, {Awan}, {Baccigalupi}, {Baildon}, {Barron}, {Battaglia}, {Battye}, {Baxter}, {Bazarko}, {Beall}, {Bean}, {Beck}, {Beckman}, {Beringue}, {Bianchini}, {Boada}, {Boettger}, {Bond}, {Borrill}, {Brown}, {Bruno}, {Bryan}, {Calabrese}, {Calafut}, {Calisse}, {Carron}, {Challinor}, {Chesmore}, {Chinone}, {Chluba}, {Cho}, {Choi}, {Coppi}, {Cothard}, {Coughlin}, {Crichton}, {Crowley}, {Crowley}, {Cukierman}, {D'Ewart}, {D{\"u}nner}, {de Haan}, {Devlin}, {Dicker}, {Didier}, {Dobbs}, {Dober}, {Duell}, {Duff}, {Duivenvoorden}, {Dunkley}, {Dusatko}, {Errard}, {Fabbian}, {Feeney}, {Ferraro}, {Flux{\`a}}, {Freese}, {Frisch}, {Frolov}, {Fuller}, {Fuzia}, {Galitzki}, {Gallardo}, {Tomas Galvez Ghersi}, {Gao}, {Gawiser}, {Gerbino}, {Gluscevic}, {Goeckner-Wald}, {Golec}, {Gordon}, {Gralla}, {Green}, {Grigorian}, {Groh}, {Groppi}, {Guan}, {Gudmundsson}, {Han}, {Hargrave}, {Hasegawa},
  {Hasselfield}, {Hattori}, {Haynes}, {Hazumi}, {He}, {Healy}, {Henderson}, {Hervias-Caimapo}, {Hill}, {Hill}, {Hilton}, {Hilton}, {Hincks}, {Hinshaw}, {Hlo{\v{z}}ek}, {Ho}, {Ho}, {Howe}, {Huang}, {Hubmayr}, {Huffenberger}, {Hughes}, {Ijjas}, {Ikape}, {Irwin}, {Jaffe}, {Jain}, {Jeong}, {Kaneko}, {Karpel}, {Katayama}, {Keating}, {Kernasovskiy}, {Keskitalo}, {Kisner}, {Kiuchi}, {Klein}, {Knowles}, {Koopman}, {Kosowsky}, {Krachmalnicoff}, {Kuenstner}, {Kuo}, {Kusaka}, {Lashner}, {Lee}, {Lee}, {Leon}, {Leung}, {Lewis}, {Li}, {Li}, {Limon}, {Linder}, {Lopez-Caraballo}, {Louis}, {Lowry}, {Lungu}, {Madhavacheril}, {Mak}, {Maldonado}, {Mani}, {Mates}, {Matsuda}, {Maurin}, {Mauskopf}, {May}, {McCallum}, {McKenney}, {McMahon}, {Meerburg}, {Meyers}, {Miller}, {Mirmelstein}, {Moodley}, {Munchmeyer}, {Munson}, {Naess}, {Nati}, {Navaroli}, {Newburgh}, {Nguyen}, {Niemack}, {Nishino}, {Orlowski-Scherer}, {Page}, {Partridge}, {Peloton}, {Perrotta}, {Piccirillo}, {Pisano}, {Poletti}, {Puddu}, {Puglisi}, {Raum}, {Reichardt},
  {Remazeilles}, {Rephaeli}, {Riechers}, {Rojas}, {Roy}, {Sadeh}, {Sakurai}, {Salatino}, {Sathyanarayana Rao}, {Schaan}, {Schmittfull}, {Sehgal}, {Seibert}, {Seljak}, {Sherwin}, {Shimon}, {Sierra}, {Sievers}, {Sikhosana}, {Silva-Feaver}, {Simon}, {Sinclair}, {Siritanasak}, {Smith}, {Smith}, {Spergel}, {Staggs}, {Stein}, {Stevens}, {Stompor}, {Suzuki}, {Tajima}, {Takakura}, {Teply}, {Thomas}, {Thorne}, {Thornton}, {Trac}, {Tsai}, {Tucker}, {Ullom}, {Vagnozzi}, {van Engelen}, {Van Lanen}, {Van Winkle}, {Vavagiakis}, {Verg{\`e}s}, {Vissers}, {Wagoner}, {Walker}, {Ward}, {Westbrook}, {Whitehorn}, {Williams}, {Williams}, {Wollack}, {Xu}, {Yu}, {Yu}, {Zago}, {Zhang}, {Zhu}, \& {Simons Observatory Collaboration}}]{2019JCAP...02..056A}
---. 2019{\natexlab{b}}, \jcap, 2019, 056, \dodoi{10.1088/1475-7516/2019/02/056}

\bibitem[{{Ajeddig} {et~al.}(2022){Ajeddig}, {André}, {Andrianasolo}, {Maury}, {Ponthieu}, {Ritacco}, \& {Shimajiri}}]{nika2pol}
{Ajeddig}, H., {André}, P., {Andrianasolo}, A., {et~al.} 2022, NIKA2pol commissioning report, Tech. rep., IRAM

\bibitem[{{Aumont} {et~al.}(2020){Aumont}, {Mac{\'\i}as-P{\'e}rez}, {Ritacco}, {Ponthieu}, \& {Mangilli}}]{2020A&A...634A.100A}
{Aumont}, J., {Mac{\'\i}as-P{\'e}rez}, J.~F., {Ritacco}, A., {Ponthieu}, N., \& {Mangilli}, A. 2020, \aap, 634, A100, \dodoi{10.1051/0004-6361/201833504}

\bibitem[{{Carroll} {et~al.}(1990){Carroll}, {Field}, \& {Jackiw}}]{1990PhRvD..41.1231C}
{Carroll}, S.~M., {Field}, G.~B., \& {Jackiw}, R. 1990, \prd, 41, 1231, \dodoi{10.1103/PhysRevD.41.1231}

\bibitem[{{Casas} {et~al.}(2021){Casas}, {Mart{\'\i}nez-Gonz{\'a}lez}, {Bermejo-Ballesteros}, {Garc{\'\i}a}, {Cubas}, {Vielva}, {Barreiro}, \& {Sanz}}]{2021Senso..21.3361C}
{Casas}, F.~J., {Mart{\'\i}nez-Gonz{\'a}lez}, E., {Bermejo-Ballesteros}, J., {et~al.} 2021, Sensors, 21, 3361, \dodoi{10.3390/s21103361}

\bibitem[{{Catalano} {et~al.}(2020){Catalano}, {Bideaud}, {Bourrion}, {Calvo}, {Fasano}, {Goupy}, {Levy-Bertrand}, {Mac{\'\i}as-P{\'e}rez}, {Ponthieu}, {Tang}, \& {Monfardini}}]{Catalano2020}
{Catalano}, A., {Bideaud}, A., {Bourrion}, O., {et~al.} 2020, \aap, 641, A179, \dodoi{10.1051/0004-6361/202038199}

\bibitem[{{Coppi} {et~al.}(2022){Coppi}, {Conenna}, {Savorgnano}, {Carrero}, {D{\"u}nner Planella}, {Galitzki}, {Nati}, \& {Zannoni}}]{coppi2022}
{Coppi}, G., {Conenna}, G., {Savorgnano}, S., {et~al.} 2022, in Society of Photo-Optical Instrumentation Engineers (SPIE) Conference Series, Vol. 12190, Millimeter, Submillimeter, and Far-Infrared Detectors and Instrumentation for Astronomy XI, ed. J.~{Zmuidzinas} \& J.-R. {Gao}, 1219015, \dodoi{10.1117/12.2628312}

\bibitem[{{Cornelison} {et~al.}(2022){Cornelison}, {Verg{\`e}s}, {Ade}, {Ahmed}, {Amiri}, {Barkats}, {Basu Thakur}, {Beck}, {Bischoff}, {Bock}, {Buza}, {Cheshire}, {Connors}, {Crumrine}, {Cukierman}, {Denison}, {Dierickx}, {Duband}, {Eiben}, {Fatigoni}, {Filippini}, {Giannakopoulos}, {Goeckner-Wald}, {Goldfinger}, {Grayson}, {Grimes}, {Hall}, {Halal}, {Halpern}, {Hand}, {Harrison}, {Henderson}, {Hildebrandt}, {Hilton}, {Hubmayr}, {Hui}, {Irwin}, {Kang}, {Karkare}, {Kefeli}, {Kovac}, {Kuo}, {Lau}, {Leitch}, {Lennox}, {Liu}, {Look}, {Megerian}, {Minutolo}, {Moncelsi}, {Nakato}, {Namikawa}, {Nguyen}, {O'brient}, {Palladino}, {Petroff}, {Prouve}, {Pryke}, {Racine}, {Reintsema}, {Salatino}, {Schillaci}, {Schmitt}, {Singari}, {Soliman}, {St. Germaine}, {Steinbach}, {Sudiwala}, {Thompson}, {Tsai}, {Tucker}, {Turner}, {Umilt{\`u}}, {Vieregg}, {Wandui}, {Weber}, {Wiebe}, {Willmert}, {Wu}, {Yang}, {Yoon}, {Young}, {Yu}, {Zeng}, {Zhang}, \& {Zhang}}]{Cornelison22}
{Cornelison}, J., {Verg{\`e}s}, C., {Ade}, P.~A.~R., {et~al.} 2022, in Society of Photo-Optical Instrumentation Engineers (SPIE) Conference Series, Vol. 12190, Millimeter, Submillimeter, and Far-Infrared Detectors and Instrumentation for Astronomy XI, ed. J.~{Zmuidzinas} \& J.-R. {Gao}, 121901X, \dodoi{10.1117/12.2620212}

\bibitem[{{Diego-Palazuelos} {et~al.}(2022){Diego-Palazuelos}, {Eskilt}, {Minami}, {Tristram}, {Sullivan}, {Banday}, {Barreiro}, {Eriksen}, {G{\'o}rski}, {Keskitalo}, {Komatsu}, {Mart{\'\i}nez-Gonz{\'a}lez}, {Scott}, {Vielva}, \& {Wehus}}]{2022PhRvL.128i1302D}
{Diego-Palazuelos}, P., {Eskilt}, J.~R., {Minami}, Y., {et~al.} 2022, \prl, 128, 091302, \dodoi{10.1103/PhysRevLett.128.091302}

\bibitem[{{Diego-Palazuelos} {et~al.}(2023){Diego-Palazuelos}, {Mart{\'\i}nez-Gonz{\'a}lez}, {Vielva}, {Barreiro}, {Tristram}, {de la Hoz}, {Eskilt}, {Minami}, {Sullivan}, {Banday}, {G{\'o}rski}, {Keskitalo}, {Komatsu}, \& {Scott}}]{2023JCAP...01..044D}
{Diego-Palazuelos}, P., {Mart{\'\i}nez-Gonz{\'a}lez}, E., {Vielva}, P., {et~al.} 2023, \jcap, 2023, 044, \dodoi{10.1088/1475-7516/2023/01/044}

\bibitem[{{D{\"u}nner} {et~al.}(2020){D{\"u}nner}, {Flux{\'a}}, {Best}, \& {Carrero}}]{Dunner2020}
{D{\"u}nner}, R., {Flux{\'a}}, J., {Best}, S., \& {Carrero}, F. 2020, in Society of Photo-Optical Instrumentation Engineers (SPIE) Conference Series, Vol. 11453, Millimeter, Submillimeter, and Far-Infrared Detectors and Instrumentation for Astronomy X, ed. J.~{Zmuidzinas} \& J.-R. {Gao}, 114532P, \dodoi{10.1117/12.2561165}

\bibitem[{Dünner {et~al.}(2021)Dünner, Fluxá, Best, Carrero, \& Boettger}]{Dunner2021}
Dünner, R., Fluxá, J., Best, S., Carrero, F., \& Boettger, D. 2021, in 2021 15th European Conference on Antennas and Propagation (EuCAP), 1--5, \dodoi{10.23919/EuCAP51087.2021.9411058}

\bibitem[{{Fasano} {et~al.}(2020){Fasano}, {Aguiar}, {Benoit}, {Bideaud}, {Bourrion}, {Calvo}, {Catalano}, {de Taoro}, {Garde}, {Gomez}, {Gomez{\^A} Renasco}, {Goupy}, {Hoarau}, {Hoyland}, {Mac{\'\i}as-P{\'e}rez}, {Marpaud}, {Monfardini}, {Pisano}, {Ponthieu}, {Rubi{\~n}o{\^A} Mart{\'\i}n}, {Tourres}, {Tucker}, {Beelen}, {Bres}, {De Petris}, {de Bernardis}, {Lagache}, {Lamagna}, {Luzzi}, {Marton}, {Masi}, {Rebolo}, \& {Roudier}}]{Fasano2020}
{Fasano}, A., {Aguiar}, M., {Benoit}, A., {et~al.} 2020, Journal of Low Temperature Physics, 199, 529, \dodoi{10.1007/s10909-019-02289-1}

\bibitem[{{Foreman-Mackey} {et~al.}(2013){Foreman-Mackey}, {Hogg}, {Lang}, \& {Goodman}}]{2013PASP..125..306F}
{Foreman-Mackey}, D., {Hogg}, D.~W., {Lang}, D., \& {Goodman}, J. 2013, \pasp, 125, 306, \dodoi{10.1086/670067}

\bibitem[{Guth(1981)}]{PhysRevD.23.347}
Guth, A.~H. 1981, Phys. Rev. D, 23, 347, \dodoi{10.1103/PhysRevD.23.347}

\bibitem[{{Hu} \& {Dodelson}(2002)}]{hu2002}
{Hu}, W., \& {Dodelson}, S. 2002, \araa, 40, 171, \dodoi{10.1146/annurev.astro.40.060401.093926}

\bibitem[{Hu \& White(1997)}]{HU1997323}
Hu, W., \& White, M. 1997, New Astronomy, 2, 323, \dodoi{https://doi.org/10.1016/S1384-1076(97)00022-5}

\bibitem[{Johnson {et~al.}(2015)Johnson, Vourch, Drysdale, Kalman, Fujikawa, Keating, \& Kaufman}]{Johnson2015}
Johnson, B.~R., Vourch, C.~J., Drysdale, T.~D., {et~al.} 2015, Journal of Astronomical Instrumentation, 04, 1550007, \dodoi{10.1142/S2251171715500075}

\bibitem[{Jost {et~al.}(2023)Jost, Errard, \& Stompor}]{PhysRevD.108.082005}
Jost, B., Errard, J., \& Stompor, R. 2023, Phys. Rev. D, 108, 082005, \dodoi{10.1103/PhysRevD.108.082005}

\bibitem[{{Kamionkowski} \& {Kovetz}(2016)}]{Kamionkowski16}
{Kamionkowski}, M., \& {Kovetz}, E.~D. 2016, \araa, 54, 227, \dodoi{10.1146/annurev-astro-081915-023433}

\bibitem[{{Keating} {et~al.}(2013){Keating}, {Shimon}, \& {Yadav}}]{Keating}
{Keating}, B.~G., {Shimon}, M., \& {Yadav}, A.~P.~S. 2013, \apjl, 762, L23, \dodoi{10.1088/2041-8205/762/2/L23}

\bibitem[{{Kovac} {et~al.}(2002){Kovac}, {Leitch}, {Pryke}, {Carlstrom}, {Halverson}, \& {Holzapfel}}]{2002Natur.420..772K}
{Kovac}, J.~M., {Leitch}, E.~M., {Pryke}, C., {et~al.} 2002, \nat, 420, 772, \dodoi{10.1038/nature01269}

\bibitem[{Linde(1982)}]{linde1982new}
Linde, A.~D. 1982, Physics Letters B, 108, 389

\bibitem[{{{LiteBIRD Collaboration}}(2023)}]{LiteBIRD:2022cnt}
{{LiteBIRD Collaboration}}. 2023, PTEP, 2023, 042F01, \dodoi{10.1093/ptep/ptac150}

\bibitem[{{Minami} \& {Komatsu}(2020)}]{2020PhRvL.125v1301M}
{Minami}, Y., \& {Komatsu}, E. 2020, \prl, 125, 221301, \dodoi{10.1103/PhysRevLett.125.221301}

\bibitem[{{Nati} {et~al.}(2017){Nati}, {Devlin}, {Gerbino}, {Johnson}, {Keating}, {Pagano}, \& {Teply}}]{Nati17}
{Nati}, F., {Devlin}, M.~J., {Gerbino}, M., {et~al.} 2017, Journal of Astronomical Instrumentation, 6, 1740008, \dodoi{10.1142/S2251171717400086}

\bibitem[{{Perotto} {et~al.}(2020){Perotto}, {Ponthieu}, {Mac{\'\i}as-P{\'e}rez}, {Adam}, {Ade}, {Andr{\'e}}, {Andrianasolo}, {Aussel}, {Beelen}, {Beno{\^\i}t}, {Berta}, {Bideaud}, {Bourrion}, {Calvo}, {Catalano}, {Comis}, {De Petris}, {D{\'e}sert}, {Doyle}, {Driessen}, {Garc{\'\i}a}, {Gomez}, {Goupy}, {John}, {K{\'e}ruzor{\'e}}, {Kramer}, {Ladjelate}, {Lagache}, {Leclercq}, {Lestrade}, {Maury}, {Mauskopf}, {Mayet}, {Monfardini}, {Navarro}, {Pe{\~n}alver}, {Pierfederici}, {Pisano}, {Rev{\'e}ret}, {Ritacco}, {Romero}, {Roussel}, {Ruppin}, {Schuster}, {Shu}, {Sievers}, {Tucker}, \& {Zylka}}]{2020A&A...637A..71P}
{Perotto}, L., {Ponthieu}, N., {Mac{\'\i}as-P{\'e}rez}, J.~F., {et~al.} 2020, \aap, 637, A71, \dodoi{10.1051/0004-6361/201936220}

\bibitem[{{Pisano} {et~al.}(2022){Pisano}, {Ritacco}, {Monfardini}, {Tucker}, {Ade}, {Shitvov}, {Benoit}, {Calvo}, {Catalano}, {Goupy}, {Leclercq}, {Macias-Perez}, {Andrianasolo}, \& {Ponthieu}}]{2022A&A...658A..24P}
{Pisano}, G., {Ritacco}, A., {Monfardini}, A., {et~al.} 2022, \aap, 658, A24, \dodoi{10.1051/0004-6361/202038643}

\bibitem[{{Planck Collaboration} {et~al.}(2020{\natexlab{a}}){Planck Collaboration}, {Aghanim, N.}, {Akrami, Y.}, {Arroja, F.}, {Ashdown, M.}, {Aumont, J.}, {Baccigalupi, C.}, {Ballardini, M.}, {Banday, A. J.}, {Barreiro, R. B.}, {Bartolo, N.}, {Basak, S.}, {Battye, R.}, {Benabed, K.}, {Bernard, J.-P.}, {Bersanelli, M.}, {Bielewicz, P.}, {Bock, J. J.}, {Bond, J. R.}, {Borrill, J.}, {Bouchet, F. R.}, {Boulanger, F.}, {Bucher, M.}, {Burigana, C.}, {Butler, R. C.}, {Calabrese, E.}, {Cardoso, J.-F.}, {Carron, J.}, {Casaponsa, B.}, {Challinor, A.}, {Chiang, H. C.}, {Colombo, L. P. L.}, {Combet, C.}, {Contreras, D.}, {Crill, B. P.}, {Cuttaia, F.}, {de Bernardis, P.}, {de Zotti, G.}, {Delabrouille, J.}, {Delouis, J.-M.}, {D\'esert, F.-X.}, {Di Valentino, E.}, {Dickinson, C.}, {Diego, J. M.}, {Donzelli, S.}, {Dor\'e, O.}, {Douspis, M.}, {Ducout, A.}, {Dupac, X.}, {Efstathiou, G.}, {Elsner, F.}, {En\ss{}lin, T. A.}, {Eriksen, H. K.}, {Falgarone, E.}, {Fantaye, Y.}, {Fergusson, J.}, {Fernandez-Cobos, R.}, {Finelli,
  F.}, {Forastieri, F.}, {Frailis, M.}, {Franceschi, E.}, {Frolov, A.}, {Galeotta, S.}, {Galli, S.}, {Ganga, K.}, {G\'enova-Santos, R. T.}, {Gerbino, M.}, {Ghosh, T.}, {Gonz\'alez-Nuevo, J.}, {G\'orski, K. M.}, {Gratton, S.}, {Gruppuso, A.}, {Gudmundsson, J. E.}, {Hamann, J.}, {Handley, W.}, {Hansen, F. K.}, {Helou, G.}, {Herranz, D.}, {Hildebrandt, S. R.}, {Hivon, E.}, {Huang, Z.}, {Jaffe, A. H.}, {Jones, W. C.}, {Karakci, A.}, {Keih\"anen, E.}, {Keskitalo, R.}, {Kiiveri, K.}, {Kim, J.}, {Kisner, T. S.}, {Knox, L.}, {Krachmalnicoff, N.}, {Kunz, M.}, {Kurki-Suonio, H.}, {Lagache, G.}, {Lamarre, J.-M.}, {Langer, M.}, {Lasenby, A.}, {Lattanzi, M.}, {Lawrence, C. R.}, {Le Jeune, M.}, {Leahy, J. P.}, {Lesgourgues, J.}, {Levrier, F.}, {Lewis, A.}, {Liguori, M.}, {Lilje, P. B.}, {Lilley, M.}, {Lindholm, V.}, {L\'opez-Caniego, M.}, {Lubin, P. M.}, {Ma, Y.-Z.}, {Mac\'{\i}as-P\'erez, J. F.}, {Maggio, G.}, {Maino, D.}, {Mandolesi, N.}, {Mangilli, A.}, {Marcos-Caballero, A.}, {Maris, M.}, {Martin, P. G.}, {Martinelli,
  M.}, {Mart\'{\i}nez-Gonz\'alez, E.}, {Matarrese, S.}, {Mauri, N.}, {McEwen, J. D.}, {Meerburg, P. D.}, {Meinhold, P. R.}, {Melchiorri, A.}, {Mennella, A.}, {Migliaccio, M.}, {Millea, M.}, {Mitra, S.}, {Miville-Desch\^enes, M.-A.}, {Molinari, D.}, {Moneti, A.}, {Montier, L.}, {Morgante, G.}, {Moss, A.}, {Mottet, S.}, {M\"unchmeyer, M.}, {Natoli, P.}, {N\o{}rgaard-Nielsen, H. U.}, {Oxborrow, C. A.}, {Pagano, L.}, {Paoletti, D.}, {Partridge, B.}, {Patanchon, G.}, {Pearson, T. J.}, {Peel, M.}, {Peiris, H. V.}, {Perrotta, F.}, {Pettorino, V.}, {Piacentini, F.}, {Polastri, L.}, {Polenta, G.}, {Puget, J.-L.}, {Rachen, J. P.}, {Reinecke, M.}, {Remazeilles, M.}, {Renault, C.}, {Renzi, A.}, {Rocha, G.}, {Rosset, C.}, {Roudier, G.}, {Rubi\~no-Mart\'{\i}n, J. A.}, {Ruiz-Granados, B.}, {Salvati, L.}, {Sandri, M.}, {Savelainen, M.}, {Scott, D.}, {Shellard, E. P. S.}, {Shiraishi, M.}, {Sirignano, C.}, {Sirri, G.}, {Spencer, L. D.}, {Sunyaev, R.}, {Suur-Uski, A.-S.}, {Tauber, J. A.}, {Tavagnacco, D.}, {Tenti, M.},
  {Terenzi, L.}, {Toffolatti, L.}, {Tomasi, M.}, {Trombetti, T.}, {Valiviita, J.}, {Van Tent, B.}, {Vibert, L.}, {Vielva, P.}, {Villa, F.}, {Vittorio, N.}, {Wandelt, B. D.}, {Wehus, I. K.}, {White, M.}, {White, S. D. M.}, {Zacchei, A.}, \& {Zonca, A.}}]{planck2018I}
{Planck Collaboration}, {Aghanim, N.}, {Akrami, Y.}, {et~al.} 2020{\natexlab{a}}, A\&A, 641, A1, \dodoi{10.1051/0004-6361/201833880}

\bibitem[{{Planck Collaboration} {et~al.}(2020{\natexlab{b}}){Planck Collaboration}, {Akrami, Y.}, {Arroja, F.}, {Ashdown, M.}, {Aumont, J.}, {Baccigalupi, C.}, {Ballardini, M.}, {Banday, A. J.}, {Barreiro, R. B.}, {Bartolo, N.}, {Basak, S.}, {Benabed, K.}, {Bernard, J.-P.}, {Bersanelli, M.}, {Bielewicz, P.}, {Bock, J. J.}, {Bond, J. R.}, {Borrill, J.}, {Bouchet, F. R.}, {Boulanger, F.}, {Bucher, M.}, {Burigana, C.}, {Butler, R. C.}, {Calabrese, E.}, {Cardoso, J.-F.}, {Carron, J.}, {Challinor, A.}, {Chiang, H. C.}, {Colombo, L. P. L.}, {Combet, C.}, {Contreras, D.}, {Crill, B. P.}, {Cuttaia, F.}, {de Bernardis, P.}, {de Zotti, G.}, {Delabrouille, J.}, {Delouis, J.-M.}, {Di Valentino, E.}, {Diego, J. M.}, {Donzelli, S.}, {Dor\'e, O.}, {Douspis, M.}, {Ducout, A.}, {Dupac, X.}, {Dusini, S.}, {Efstathiou, G.}, {Elsner, F.}, {En\ss{}lin, T. A.}, {Eriksen, H. K.}, {Fantaye, Y.}, {Fergusson, J.}, {Fernandez-Cobos, R.}, {Finelli, F.}, {Forastieri, F.}, {Frailis, M.}, {Franceschi, E.}, {Frolov, A.}, {Galeotta, S.},
  {Galli, S.}, {Ganga, K.}, {Gauthier, C.}, {G\'enova-Santos, R. T.}, {Gerbino, M.}, {Ghosh, T.}, {Gonz\'alez-Nuevo, J.}, {G\'orski, K. M.}, {Gratton, S.}, {Gruppuso, A.}, {Gudmundsson, J. E.}, {Hamann, J.}, {Handley, W.}, {Hansen, F. K.}, {Herranz, D.}, {Hivon, E.}, {Hooper, D. C.}, {Huang, Z.}, {Jaffe, A. H.}, {Jones, W. C.}, {Keih\"anen, E.}, {Keskitalo, R.}, {Kiiveri, K.}, {Kim, J.}, {Kisner, T. S.}, {Krachmalnicoff, N.}, {Kunz, M.}, {Kurki-Suonio, H.}, {Lagache, G.}, {Lamarre, J.-M.}, {Lasenby, A.}, {Lattanzi, M.}, {Lawrence, C. R.}, {Le Jeune, M.}, {Lesgourgues, J.}, {Levrier, F.}, {Lewis, A.}, {Liguori, M.}, {Lilje, P. B.}, {Lindholm, V.}, {L\'opez-Caniego, M.}, {Lubin, P. M.}, {Ma, Y.-Z.}, {Mac\'{\i}as-P\'erez, J. F.}, {Maggio, G.}, {Maino, D.}, {Mandolesi, N.}, {Mangilli, A.}, {Marcos-Caballero, A.}, {Maris, M.}, {Martin, P. G.}, {Mart\'{\i}nez-Gonz\'alez, E.}, {Matarrese, S.}, {Mauri, N.}, {McEwen, J. D.}, {Meerburg, P. D.}, {Meinhold, P. R.}, {Melchiorri, A.}, {Mennella, A.}, {Migliaccio, M.},
  {Mitra, S.}, {Miville-Desch\^enes, M.-A.}, {Molinari, D.}, {Moneti, A.}, {Montier, L.}, {Morgante, G.}, {Moss, A.}, {M\"unchmeyer, M.}, {Natoli, P.}, {N\o{}rgaard-Nielsen, H. U.}, {Pagano, L.}, {Paoletti, D.}, {Partridge, B.}, {Patanchon, G.}, {Peiris, H. V.}, {Perrotta, F.}, {Pettorino, V.}, {Piacentini, F.}, {Polastri, L.}, {Polenta, G.}, {Puget, J.-L.}, {Rachen, J. P.}, {Reinecke, M.}, {Remazeilles, M.}, {Renzi, A.}, {Rocha, G.}, {Rosset, C.}, {Roudier, G.}, {Rubi\~no-Mart\'{\i}n, J. A.}, {Ruiz-Granados, B.}, {Salvati, L.}, {Sandri, M.}, {Savelainen, M.}, {Scott, D.}, {Shellard, E. P. S.}, {Shiraishi, M.}, {Sirignano, C.}, {Sirri, G.}, {Spencer, L. D.}, {Sunyaev, R.}, {Suur-Uski, A.-S.}, {Tauber, J. A.}, {Tavagnacco, D.}, {Tenti, M.}, {Toffolatti, L.}, {Tomasi, M.}, {Trombetti, T.}, {Valiviita, J.}, {Van Tent, B.}, {Vielva, P.}, {Villa, F.}, {Vittorio, N.}, {Wandelt, B. D.}, {Wehus, I. K.}, {White, S. D. M.}, {Zacchei, A.}, {Zibin, J. P.}, \& {Zonca, A.}}]{Planck2018X}
{Planck Collaboration}, {Akrami, Y.}, {Arroja, F.}, {et~al.} 2020{\natexlab{b}}, A\&A, 641, A10, \dodoi{10.1051/0004-6361/201833887}

\bibitem[{Polnarev(1985)}]{polnarev1985polarization}
Polnarev, A. 1985, Soviet Astronomy, 29, 607

\bibitem[{Pospelov {et~al.}(2009)Pospelov, Ritz, \& Skordis}]{PhysRevLett.103.051302}
Pospelov, M., Ritz, A., \& Skordis, C. 2009, Phys. Rev. Lett., 103, 051302, \dodoi{10.1103/PhysRevLett.103.051302}

\bibitem[{Qin {et~al.}(2020)Qin, Poulin, Mesinger, Greig, Murray, \& Park}]{10.1093/mnras/staa2797}
Qin, Y., Poulin, V., Mesinger, A., {et~al.} 2020, Monthly Notices of the Royal Astronomical Society, 499, 550, \dodoi{10.1093/mnras/staa2797}

\bibitem[{{Rees}(1968)}]{1968ApJ...153L...1R}
{Rees}, M.~J. 1968, \apjl, 153, L1, \dodoi{10.1086/180208}

\bibitem[{{Ritacco} {et~al.}(2023){Ritacco}, {Boulanger}, {Guillet}, {Delouis}, {Puget}, {Aumont}, \& {Vacher}}]{2023A&A...670A.163R}
{Ritacco}, A., {Boulanger}, F., {Guillet}, V., {et~al.} 2023, \aap, 670, A163, \dodoi{10.1051/0004-6361/202244269}

\bibitem[{{Ritacco} {et~al.}(2017){Ritacco}, {Ponthieu}, {Catalano}, {Adam}, {Ade}, {Andr{\'e}}, {Beelen}, {Beno{\^\i}t}, {Bideaud}, {Billot}, {Bourrion}, {Calvo}, {Coiffard}, {Comis}, {D{\'e}sert}, {Doyle}, {Goupy}, {Kramer}, {Leclercq}, {Mac{\'\i}as-P{\'e}rez}, {Mauskopf}, {Maury}, {Mayet}, {Monfardini}, {Pajot}, {Pascale}, {Perotto}, {Pisano}, {Rebolo-Iglesias}, {Rev{\'e}ret}, {Rodriguez}, {Romero}, {Ruppin}, {Savini}, {Schuster}, {Sievers}, {Thum}, {Triqueneaux}, {Tucker}, \& {Zylka}}]{2017A&A...599A..34R}
{Ritacco}, A., {Ponthieu}, N., {Catalano}, A., {et~al.} 2017, \aap, 599, A34, \dodoi{10.1051/0004-6361/201629666}

\bibitem[{{Ritacco} {et~al.}(2018){Ritacco}, {Mac{\'\i}as-P{\'e}rez}, {Ponthieu}, {Adam}, {Ade}, {Andr{\'e}}, {Aumont}, {Beelen}, {Beno{\^\i}t}, {Bideaud}, {Billot}, {Bourrion}, {Bracco}, {Calvo}, {Catalano}, {Coiffard}, {Comis}, {D'Addabbo}, {De Petris}, {D{\'e}sert}, {Doyle}, {Goupy}, {Kramer}, {Lagache}, {Leclercq}, {Lestrade}, {Mauskopf}, {Mayet}, {Maury}, {Monfardini}, {Pajot}, {Pascale}, {Perotto}, {Pisano}, {Rebolo-Iglesias}, {Rev{\'e}ret}, {Rodriguez}, {Romero}, {Roussel}, {Ruppin}, {Schuster}, {Sievers}, {Siringo}, {Thum}, {Triqueneaux}, {Tucker}, {Wiesemeyer}, \& {Zylka}}]{2018A&A...616A..35R}
{Ritacco}, A., {Mac{\'\i}as-P{\'e}rez}, J.~F., {Ponthieu}, N., {et~al.} 2018, \aap, 616, A35, \dodoi{10.1051/0004-6361/201731551}

\bibitem[{{Ritacco} {et~al.}(2022){Ritacco}, {Adam}, {Ade}, {Ajeddig}, {Andr{\'e}}, {Artis}, {Aumont}, {Aussel}, {Beelen}, {Beno{\^\i}t}, {Berta}, {Bing}, {Bourrion}, {Calvo}, {Catalano}, {De Petris}, {D{\'e}sert}, {Doyle}, {Driessen}, {Gomez}, {Goupy}, {K{\'e}ruzor{\'e}}, {Kramer}, {Ladjelate}, {Lagache}, {Leclercq}, {Lestrade}, {Mac{\'\i}as-P{\'e}rez}, {Maury}, {Mauskopf}, {Mayet}, {Monfardini}, {Mu{\~n}oz-Echeverr{\'\i}a}, {Perotto}, {Pisano}, {Ponthieu}, {Rev{\'e}ret}, {Rigby}, {Romero}, {Roussel}, {Ruppin}, {Schuster}, {Shu}, {Sievers}, {Tucker}, \& {Zylka}}]{2022EPJWC.25700042R}
{Ritacco}, A., {Adam}, R., {Ade}, P., {et~al.} 2022, in European Physical Journal Web of Conferences, Vol. 257, mm Universe @ NIKA2 - Observing the mm Universe with the NIKA2 Camera, 00042, \dodoi{10.1051/epjconf/202225700042}

\bibitem[{{Rosset, C.} {et~al.}(2010){Rosset, C.}, {Tristram, M.}, {Ponthieu, N.}, {Ade, P.}, {Aumont, J.}, {Catalano, A.}, {Conversi, L.}, {Couchot, F.}, {Crill, B. P.}, {Désert, F.-X.}, {Ganga, K.}, {Giard, M.}, {Giraud-Héraud, Y.}, {Haïssinski, J.}, {Henrot-Versillé, S.}, {Holmes, W.}, {Jones, W. C.}, {Lamarre, J.-M.}, {Lange, A.}, {Leroy, C.}, {Macías-Pérez, J.}, {Maffei, B.}, {de Marcillac, P.}, {Miville-Deschênes, M.-A.}, {Montier, L.}, {Noviello, F.}, {Pajot, F.}, {Perdereau, O.}, {Piacentini, F.}, {Piat, M.}, {Plaszczynski, S.}, {Pointecouteau, E.}, {Puget, J.-L.}, {Ristorcelli, I.}, {Savini, G.}, {Sudiwala, R.}, {Veneziani, M.}, \& {Yvon, D.}}]{Rosset2010}
{Rosset, C.}, {Tristram, M.}, {Ponthieu, N.}, {et~al.} 2010, A\&A, 520, A13, \dodoi{10.1051/0004-6361/200913054}

\bibitem[{Treuttel {et~al.}(2023)Treuttel, Gatilova, Caroopen, Feret, Gay, Vacelet, Valentin, Jin, Cavanna, Jacob, Mignoni, Lavignolle, Krieg, Goldstein, Courtade, Larigauderie, Ravanbakhsh, Garcia, Maestrini, \& Hartogh}]{Treuttel}
Treuttel, J., Gatilova, L., Caroopen, S., {et~al.} 2023, IEEE Transactions on Terahertz Science and Technology, 13, 324, \dodoi{10.1109/TTHZ.2023.3263623}

\bibitem[{{Vacher} {et~al.}(2023){Vacher}, {Aumont}, {Boulanger}, {Montier}, {Guillet}, {Ritacco}, \& {Chluba}}]{2023A&A...672A.146V}
{Vacher}, L., {Aumont}, J., {Boulanger}, F., {et~al.} 2023, \aap, 672, A146, \dodoi{10.1051/0004-6361/202245292}

\bibitem[{{Verg{\`e}s} {et~al.}(2021){Verg{\`e}s}, {Errard}, \& {Stompor}}]{2021PhRvD.103f3507V}
{Verg{\`e}s}, C., {Errard}, J., \& {Stompor}, R. 2021, \prd, 103, 063507, \dodoi{10.1103/PhysRevD.103.063507}

\end{thebibliography}
\bibliographystyle{aasjournal}



\end{document}